  \documentclass[utf8]{FrontiersinHarvard}

\newcommand{\myskip}[1]{}

\DeclareMathAlphabet{\mathpzc}{OT1}{pzc}{m}{it}

\newcommand{\specl}{{\rm spec}_l}

\newcommand{\hnpa}{{(n,\alp)}}
\newcommand{\hnma}{{(n,\malp)}}

\newcommand{\hspmt}{\hspace{-10mm}}

\newcommand{\GN}{{G_N}}
\newcommand{\diag}{{\rm diag}}
\newcommand{\dyn}{{\rm dyn}}

\newcommand{\co}{{\rm co}} 
\newcommand{\si}{{\rm si}}

\newcommand{\qad}{\hspace{1mm}}
\newcommand{\iss}{{\hspace{0.2mm}=\hspace{0.2mm}}}

\newcommand{\pplus}{{\hspace{-0.005mm} + \hspace{-0.005mm}}}
\newcommand{\mmin}{{\hspace{-0.005mm} - \hspace{-0.005mm}}}

\newcommand{\sign}{\sig_n}

\newcommand{\sigi}{\sig_i}

\newcommand{\met}{{\bf m}}

\newcommand{\hm}{{\hat m}}
\newcommand{\sig}{\sigma}

\newcommand{\hsig}{\hat\sigma}

\newcommand{\hi}{{(i)}}
\newcommand{\hen}{{(n)}}
\newcommand{\heN}{{(N)}}

\newcommand{\alp}{\alpha}
\newcommand{\malp}{{-\alpha}}
\newcommand{\bet}{\beta}
\newcommand{\mbet}{{-\beta}}
\newcommand{\gam}{\gamma}

\newcommand{\mhalf}{{-\half}}

\renewcommand{\xi}{\delta N}
\newcommand{\henn}{{(n)}}

\newcommand{\Fdyns}{F_\dyn ^s}
\newcommand{ \dotFdyns}{\dot F_\dyn ^s}

\renewcommand{\sb}{\bar s}
\newcommand{\heen}{{(1)}}
\newcommand{\htwee}{{(2)}}
\newcommand{\hN}{{(N)}}

\newcommand{\dc}{{\rm dc}}

\newcommand{\reg}{{\rm reg}}

\newcommand{\pama}{{\rm pm}}

\newcommand{\half}{\frac{1}{2}}

\newcommand{\eps}{\varepsilon}
\renewcommand{\d}{{\rm d}}

\newcommand{\rmG}{{\rm G}}
\newcommand{\rmM}{{\rm M}}

\newcommand{\vm}{{\bf m}}
\newcommand{\om}{\omega}
\newcommand{\Om}{\Omega}

\newcommand{\ti}{t_{\rm i}}

\newcommand{\cF}{F}

\newcommand{\SA}{{\rm SA}}

\newcommand{\BEQ}{\begin{eqnarray}}
\newcommand{\EEQ}{\end{eqnarray}}
\newcommand{\BEA}{\begin{eqnarray}}
\newcommand{\EEA}{\end{eqnarray}}
\newcommand{\nn}{\nonumber}

\renewcommand{\sb}{{\bar s} }

\def\keyFont{\fontsize{8}{11}\helveticabold }
\def\firstAuthorLast{T.M. Nieuwenhuizen} 
\def\Authors{Theodorus Maria Nieuwenhuizen}
\def\Address{Institute for Theoretical Physics,  University of Amsterdam 
 \\ Science Park 904, 1090 GL  Amsterdam, The Netherlands} 
\def\corrAuthor{T.M. Nieuwenhuizen}

\def\corrEmail{t.m.nieuwenhuizen@uva.nl}

\myskip{ 
} 

\begin{document}
 \onecolumn
 

\title{Dynamics of  ideal quantum measurement of a  spin 1 with a Curie-Weiss magnet }  

\myskip{}{
\author[\firstAuthorLast ]{\Authors} 
\address{} 
\correspondance{} 

 \extraAuth{}

 \maketitle

\begin{abstract}
\section{}
Quantum measurement is a dynamical process of an apparatus coupled to a test system.
Ideal measurement of the $z$-component of a spin-$\half$ ($s_z=\pm\frac{1}{2}$) has been modeled by the Curie-Weiss model for quantum measurement.
Recently, the model was generalized to higher spin and the thermodynamics was solved. Here the dynamics is considered.
To this end, the dynamics for spin-$\half$ case are cast in general notation.
The dynamics of the measurement of the $z$-component of a spin-1 ($s_z=0,\pm 1$) are solved in detail and evaluated numerically.
Energy costs of the measurement, which are macroscopic, are evaluated.
Generalization to higher spin is straightforward. 
\tiny
 \keyFont{ \section{Keywords:} ideal quantum measurement, dynamics, Curie-Weiss model, higher spin, exact solution}
\end{abstract}
} 


\section*{Table of contents}
\noindent\noindent
\ref{sec1} \hspace{2mm} Introduction \\
\ref{sec1.1} \hspace{2mm} Higher spin models \\
\ref{sec2}   \hspace{2mm}  Higher-spin  Curie-Weiss Hamiltonian models \\
\ref{sec2.1}  \hspace{2mm}   Spin-spin Hamiltonian of the magnet  \\
\ref{sec2.2} \hspace{2mm} The interaction Hamiltonian \\
\ref{sec2.3} \hspace{2mm}  Coupling to a harmonic oscillator bath \\
\ref{sec2.4} \hspace{2mm}  Evolution of the density matrix  \\
\ref{sec2.5} \hspace{2mm}  On physical implementation \\
\ref{sec3} \hspace{2mm} The spin $\frac{1}{2}$ case reformulated\\
\ref{sec3.1}  \hspace{2mm} Elements of the statics \\
\ref{sec3.2} \hspace{2mm}  The interaction Hamiltonian \\
\ref{sec4} \hspace{2mm}  Dynamics of the spin $\half$ model \\
\ref{sec4.1} \hspace{2mm} Truncation for spin $\half$ \\
\ref{sec4.2} \hspace{2mm}  Registration for spin $\half$ \\
\ref{sec4.3} \hspace{2mm}  $H$-theorem and relaxation to equilibrium  \\
\ref{sec4.4} \hspace{2mm} Decoupling the apparatus \\
\ref{sec5} \hspace{2mm} Dynamics of the spin 1 model \\
\ref{sec5.1} \hspace{2mm} Off-diagonal sector: truncation of cat terms \\
\ref{sec5.1.1} \hspace{2mm} Initial regime: Dephasing \\
\ref{sec5.1.2} \hspace{2mm} Follow up: Decoherence \\
\ref{sec5.2} \hspace{2mm} Registration dynamics for spin 1 \\
\ref{sec5.3} \hspace{2mm} $H$-theorem and relaxation to equilibrium \\
\ref{sec5.4} \hspace{2mm} Numerical analysis  \\
\ref{sec5.5} \hspace{2mm} Decoupling the apparatus \\
\ref{sec5.6} \hspace{2mm} Energy cost of quantum measurement \\
\ref{sec6}    \hspace{2mm}  Summary \\
\ref{refs} \hspace{2mm}  References

\section{Introduction}
\label{sec1}

This year we celebrate the centennial of the formulation of quantum theory, see \cite{capellmann2017quantum} for  prehistory.
After the ``Zur Quantummechanik" by \cite{born1925quantenmechanik}, the Dreim\"anner  Arbeit by Born, Heisenberg and Jordan (\cite{born1926quantenmechanik})
on the matrix mechanics  was soon followed by the \cite{schrodinger1926undulatory} formulation of wave mechanics, inspired by insights of  \cite{de1924recherches}.
The predictive power of the theory was expressed by the \cite{born1926bquantenmechanik} rule. 
For a compilation of historic contributions, see \cite{wheeler2014quantum}.

The interpretation of quantum mechanics has been discussed during the whole century since then. 
The Copenhagen interpretation -- the Born rule and the collapse postulate -- emerged as the most reasonable one. 
Many attempts to deepen the understanding start from these postulates. However, they are merely shortcuts for what happens in a laboratory.
With our collaborators Armen Allahverdyan and Roger Balian, we have taken the viewpoint to start from the uninterpreted quantum formalism and employ it
for the dynamics of an idealized measurement. Whatever is solved already in this approach, needs not be interpreted; interpretation is needed to 
put the results in a proper, global context.
 As discussed below, this effort  has led to a specified version of the statistical interpretation of quantum mechanics, popularized by \cite{ballentine1970statistical}.

The present paper deals with the dynamics of an ideal quantum measurement. 
It is based on the Curie-Weiss model for measuring the $z$-component of a spin $\half$, introduced in  \cite{allahverdyan2003curie}. 
After reviewing various models for quantum measurement, it is  considered in great detail in 2013 \cite{allahverdyan2013understanding}.
The apparatus consists of a mean-field type magnet having $N\gg1$ spins $\half$, coupled to a harmonic oscillator bath. 
The magnet starts in a metastable, long-lived, paramagnetic state, separated by free energy barriers from the stable states with upward or downward magnetization.
It is in a ``ready'' state for use in a measurement.

When employed as an apparatus, the magnetization acts as a pointer for the outcome. 
The coupling to the tested spin causes a quick transition to one of the stable states, thereby registering the measurement.
For this to succeed, the coupling must be large enough to overcome the free energy barrier.
While the final state of the magnet is described by thermodynamics, much detail is contained in the dynamical evolution towards this state.

In an ideal measurement, the Born rule appears due to the non-disturbance of the measured operator.
It provides the probabilities for the  pointer, that is, for the final magnetization to be upward or downward.
The state of the microscopic spin is correlated to it, and inferred from the pointer indication. 

Understanding the dynamics also provides a natural route towards interpretation of quantum mechanics.
Indeed, when assuming the quantum formalism, the task is to work out its predictions and only then to interpret the results.
 This leads to view the wave function, or more generally: the density matrix, 
as a state of best knowledge, and the ``collapse of the wave function'' or ``disappearance of cat states'' as an update of knowledge after selection of the 
runs with identical outcomes, compatible with the quantum formalism. 
Notably, quantum theory is not a theory of Nature based somehow on an ontology, rather it is an abstract construct to explain its probabilistic features. 

The ``measurement problem'', that is, to describe  the individual experiments that occur in a laboratory, is in our view still the most outstanding challenge of modern science.
Many attempts have been made to solve it by making adaptations or small alterations of quantum mechanics, or by 
interpreting it in a different way.
We hold the opinion that this whole enterprise is in vain; one should start completely from scratch, in order to
``derive quantum mechanics'', that is to say, establish the origin of quantum behavior in Nature \footnote{{An 
analogy is offered by the dark matter problem in cosmology. 
Abandoning particle dark matter, we view dark ``matter'' as a form of energy, 
and assume new properties of vacuum energy. This provides a description of black holes with a core rather than a singularity \citep{nieuwenhuizen2023exact};
aspects of dark matter throughout the history and future of the Universe \citep{nieuwenhuizen2024solution};  
and the giant dark matter clouds around isolated galaxies \citep{nieuwenhuizen2024indefinitely}, explaining the ``indefinite flattening'' 
of their rotation curves \citep {mistele2024indefinitely}. Remarkably, this approach is a generalization of the classical 
Lorentz-Poincar\'e electron -- a charged, non-spinning spherical shell filled with vacuum energy \citep{nieuwenhuizen2025how}.
}}

Various formalisms of quantum mechanics are reviewed by \cite{david2015formalisms}.
The insight that quantum mechanics is only meaningful in a laboratory context, stressed in particular by Bohr,
 is central in the approach of  \cite{auffeves2016contexts,auffeves2020deriving} and leads to new insights for the Heisenberg cut between quantum and classical
\citep{van2023contextual}.
One century of interpretation of the Born rule, including the modern one, is overviewed by \cite{neumaier2025born}.

\subsection{The Curie-Weiss model for quantum measurement}

A macroscopic material consists of atoms, which are quantum particles. The starting point for their dynamics
lies in quantum statistical mechanics. For a measurement,  the apparatus must be macroscopic and have a macroscopic pointer,
so that the outcome of the measurement can be read off or processed automatically. Hereto an operator formalism is required,
with dynamics set by the Liouville-von Neumann equation, the generalization of the Schr\"odinger equation to mixed states.

Progress on solvable  models for quantum meaasurement was made in last decades, when we, together with A. Allahverdyan and R. Balian in our ``ABN''  collaboration, 
introduced and solved the so-called Curie-Weiss model for quantum measurement \cite{allahverdyan2003curie}. 
Here the classical Curie-Weiss model of a magnet is taken in its quantum version and applied to the measurement of a quantum spin $\half$.
 Various further aspects were 
presented in \cite{allahverdyan2003quantuma,allahverdyan2005quantum,allahverdyan2005dynamics,allahverdyan2007quantum,allahverdyan2006phase}.
They were reviewed and greatly expanded in  \cite{allahverdyan2013understanding}.
Lecture notes were presented by \cite{nieuwenhuizen2014lectures}.
A straightforward interpretation for a class of such measurements models was provided by \cite{allahverdyan2017sub};
it is a specified version of the statistical interpretation, made popular by \cite{ballentine1970statistical}.

Simultaneous measurement of two noncommuting quantum variables was worked out \citep{perarnau2017simultaneous}, 
as well as an application to Einstein-Podolsky-Rosen type of measurements \citep{perarnau2017dynamics}.
A numerical test on a simplified version of the Curie-Weiss model reproduced nearly all of its properties \citep{donker2018quantum}.

Our ensuing insights suitable for teachers of quantum theory (at high school, bachelor or master level) are presented in  \cite{allahverdyan2024teaching} 
and summarized in a features article \citep{allahverdyan2025interpretation}.

\subsection{Higher spin Curie-Weiss models}
\label{sec1.1}

The mentioned Curie-Weiss  model is recently generalized by us to measure a spin $l > \half$ \citep{nieuwenhuizen2022models}, 
a paper to be termed ``Models'' henceforth. For spin $l$,  the state of the magnet is described by $2l$ order parameters.
To assure an unbiased measurement, the Hamiltonian of the apparatus and the interaction Hamiltonian with the tested system, have $Z_{2l+1}$ symmetry.
The statics was solved for spin 1, $\frac{3}{2}$, 2 and $\frac{5}{2}$. 

Here the dynamics is worked out for spin 1, laying the groundwork for higher spin dynamics.
In the spin $\half$ Curie Weiss model it was found that Schr\"odinger cat terms disappear by two mechanisms, 
dephasing of the magnet, possibly followed by decoherence due to the thermal bath.
Similar behavior is now investigated for spin $1$.

The setup of the paper is as follows.
In section \ref{sec2} we recall the formulation of the Curie-Weiss model for general spin-$l$
and discuss aspects of its physical implementation for spin $\half$ and spin 1.
In section \ref{sec3} we revisit the spin-$\half$ case  and cast its dynamics in general form.
In section \ref{sec5} we  analyze the dynamics of the spin-$1$ situation.
We close with a  summary, section \ref{summ}.

\renewcommand{\thesection}{\arabic{section}}
\section{Higher-spin  Curie-Weiss Hamiltonian models}
\renewcommand{\theequation}{2.\arabic{equation}}
\setcounter{equation}{0} 
\renewcommand{\thesection}{\arabic{section}}

\label{sec2}

We start with recalling  some properties of higher spin models, that we introduced in ``Models''  \citep{nieuwenhuizen2022models}.
The statics was considered there; here we define and study the dynamics,  recall parts of the spin $\half$ case.
We often refer to the review by \cite{allahverdyan2013understanding}, to be termed ``Opus''.

In the following, we denote quantum operators by a hat, specifically $\hat {\bf s}$ and $\hat s_z$ for the measured spin, and 
{\boldmath{$\hat \sigma^\hi$}} and $\hsig_z^\hi$ for the spins of the apparatus.  For simplicity of notation, we follow Models and 
denote the eigenvalues  without a hat, notably those of $\hat s_z$ by $s$ and the ones of $\hsig_z^\hi$ by $\sig_i$.
Sums over $i$ lead to the operators $\hat m_k$ and their scalar values $m_k$ for $k=1,2,\cdots,2l$.
Switching between these operators and their eigenvalues is straightforward.

The strategy is to measure the $z$-component of a quantum spin-$l$ with ($l=\frac{1}{2},1,\frac{3}{2},\cdots$).
The eigenvalues $s$ of the operator $\hat s_z$  lie in the spectrum\footnote{To simplify the notation, 
we replace the standard notation for spins by $s\to l$ and $s_z\to s$. 
For an angular momentum $ L^2=l(l+1)$, the model also applies to the measurement of  $\hat L_z$
with eigenvalues $m\to s$. We employ units $\hbar=k=1$.}
\BEQ \label{specl}
s \in \specl=\{-l,-l+1,\cdots,l-1,l \} .
\EEQ
The measurement will be performed by employing an apparatus with $N\gg1$ vector spins-$l$ having operators  {\boldmath{$\hat \sigma^\hi$}},
$i=1,\cdots,N$. 
They have components $\hat\sigma^\hi_{a}$ ($a=x,y,z$), with eigenvalues $\sigma^\hi_{a}  \in \specl$. 
These operators are mutually coupled in the Hamiltonian of M.
For each $i=1,\cdots N$, and for each $\hat\sigma^\hi_{a}$, $a=x,y,z$ they are also coupled to a thermal harmonic oscillator bath; 
for the case $l=\half$ this was worked out \cite{allahverdyan2003curie,allahverdyan2013understanding}.  
The generalization of such a bath for arbitrary spin-$l$ is straightforward and will be applied to the spin-1 model.

\subsection{Spin-spin Hamiltonian of the magnet}
\label{sec2.1}

A  quantum measurement is often supposed to be ``instantaneous''. In our idealized modeling, it will take a finite time, but the tested spin
will not evolve in mean time. In other words,  the spin itself  is ``sitting still'' and  waiting to be measured.
Neither should it evolve during the  ``fast'' measurement.
This is realized when its Hamiltonian $\hat H_{\rm S}$ commutes with $\hat s_z$; 
we consider the simplest case:  $\hat H_{\rm S}=0$.

In order to have an unbiased apparatus, the Hamiltonian of the magnet should have degenerate minima and maximal symmetry.
To construct such a functional,  we consider,  in the eigenvalue presentation,  the form
\BEQ \label{C2=}
C_2=\nu^2\sum_{i,j=1}^N \cos\frac{2\pi (\sigma_i-\sigma_j) }{2l+1} ,\quad \nu\equiv\frac{1}{N},
\EEQ
which is maximal in the ferromagnetic states $\sig_i=\sig_1$ ($i=2,\cdots,N$).
In general,  such  interactions do not seem realistic, but here the cosine rule allows to express this as spin-spin interactions,
\BEQ \label{C2a=}
C_2
=\co_l^2+\si_l^2 ,
\EEQ
which is bilinear in the single-spin sums
\BEQ
\co_l \iss \frac{1}{N}\sum_{i=1}^N \cos\frac{2\pi\sigma_i }{2l \pplus 1} , \qad
\si_l \iss \frac{1}{N}\sum_{i=1}^N \sin\frac{2\pi\sigma_i }{2l \pplus 1}   .
\EEQ
The discrete values of the spin projections
allow to express these terms in  the $2l$ spin moments,
\BEQ \label{mk=}
m_k=\frac{1}{N}\sum_{i=1}^N\sigma_i^k,\qquad (k=1,\cdots,2l) ,
\EEQ
while $m_0\equiv 1$.
For $l=\half$, the values $s=\pm\half$  imply
\BEQ
\cos\pi s=0 ,\quad \sin \pi s=2s.
\EEQ
Applying this for $s\to\sig_i$ and summing over $i$, yields
\BEQ
\co_{\half}=0,\quad
\si_{\half}=2m_1,\quad m_1=\frac{1}{N}\sum_{i=1}^N \sig_i .
\EEQ

 In the case $l=1$, one has $s=0,\pm1$. The rule
\BEQ \label{exps2s}
\cos\frac{2\pi s}{3}=1-\frac{3}{2}s^2 ,
\quad
\sin\frac{2\pi s}{3}=  \frac{\sqrt{3}}{2}s ,
\EEQ
leads for $s\to\sig_i$ and summing over $i$, to \BEQ
\co_{1}=1-\frac{3}{2}m_2,\quad
\si_{1}=\frac{\sqrt{3}}{2}m_1,
\EEQ
Here $m_2$ ranges from 0 to 1 with steps of $\nu\equiv 1/N$,  while $m_1$ ranges from $-m_2$ to $m_2$ with steps of $2\nu$.
At finite $N$ one can label the discrete $m_{1,2}$ as
\BEQ &&
m_1=(2n_1-n_2)\nu, \quad  m_2=n_2\nu,\quad  \\ && \nn (0\le n_2\le N,\quad  0\le n_1\le n_2) .
\EEQ
Results for $s=\frac{3}{2}$, $2$ and $\frac{5}{2}$ are given in Models.

Let out of the $N$ spins $\sigma_i$,  a number $N_\sigma=\sum_i\delta_{\sigma_i, \, \sigma}$
 take the value  $\sigma\in \specl$ and let $x_\sigma =N_\sigma /N$ be their fraction.
The sum rule $\sum_{\sigma}N_\sigma =N$ implies $m_0\equiv \sum_{\sigma}x_\sigma =1$. The moments read 
\BEQ\label{mk=}
m_k=\sum_{\sigma=-l}^l x_\sigma \sigma^k,\quad k=1,\cdots,2l,\qquad  
\EEQ
Inversion of these relations determines the $x_\sigma$ as linear combinations of the $m_k$.
For $l=\half$ one has
\BEQ
m_1=\half x_\half-\half x_{-\half},\quad 
x_{\pm\half}=\half\pm m_1 .  
\EEQ

For spin 1 ($l=1$) one has 
\BEQ \label{msl=1}
m_1=-x_{-1}+x_1,\qquad m_2=x_{-1}+x_1 .
\EEQ
With $x_{-1}+x_0+x_1=1$ their inversion reads
 \BEQ\label{xsl=1}
 x_0=1-m_2,\qquad x_{\pm 1}=\frac{m_2\pm m_1}{2} .
\EEQ

In a quantum approach, one goes to operators,  and sets $s\to\hat s_z$,  $\sigma_i\to\hat \sigma_z^\hi$ and  $m_k \to \hat m_k$.
For the Hamiltonian $\hat H_\rmM =N\hat H$ we  follow \cite{allahverdyan2003curie} and adopt the spin-spin and four-spin interactions
\BEQ\label{UN=}
\hat H_\rmM =N \hat H,\quad 
\hat H=-\frac{1}{2}J_2 \hat C_2-\frac{1}{4}J_4 \hat C_2^2 .
\EEQ
Multi-spin interaction terms like $-\frac{1}{6}J_6\hat C_2^3-\frac{1}{8}J_8\hat  C_2^4$ can be added,
without changing the overall picture.

\subsection{The interaction Hamiltonian}
\label{sec2.2}

The coupling between the tested spin S and the magnet M
is chosen similar to (\ref{C2=}), 
\BEQ\label{HSAs} 
\hspace{-2mm}
\hat H_\SA =N \hat I,\qquad 
\hat I= \mmin \frac{g}{N}\sum_{i=1}^N \cos\frac{2\pi (\hat s_z \mmin \hat \sigma_z^\hi )}{2l+1} ,
\EEQ
where $g$ is the coupling constant.
It takes the values
\BEQ   \hspmt
I_s(\{\sig_i\}) = -\frac{g}{N}\sum_{i=1}^N  \big( 
\cos\frac{2\pi s}{2l+1} \cos\frac{2\pi \sigma_i}{2l+1} 
+\sin\frac{2\pi s}{2l+1} \sin\frac{2\pi \sigma_i}{2l+1} \big) ,
\EEQ 
This can be expressed as a linear combination of the moments $m_1$, $\cdots$, $m_{2l}$.
For $l=\half$ one has
\BEQ \label{HIcsh}
I_s(m_1)=-4gsm_1,
\EEQ
and for $l=1$,  denoting $\vm=(m_1,m_2)$,
\BEQ
I_s(\vm)=-g \big [ (1 \mmin \frac{3}{2}s^2 )(1 \mmin \frac{3}{2} m_2) \pplus \frac{3}{4}s m_1 \big] .
\EEQ

The total spin Hamiltonian, 
\BEQ
\hat H=\hat H_\rmM +\hat H_\SA  = \hat H_\rmM - N\hat I ,
\EEQ
has $Z_{2l+1}$ symmetry: on the diagonal basis, a shift $s\to s+\sb$ with $\sb=1,2,\cdots,2l+1$ can be accompanied by a shift
$\sig_i\to \sig_i+\sb$ for all $i$. This is evident in the cosine expressions, and implies a somewhat hidden invariance in the 
formulation in terms of the moments $m_k$, as discussed in Models.

\subsection{Coupling to a harmonic oscillator bath}
\label{sec2.3}

For general spin $l$, the magnet-bath coupling is taken as the spin-boson coupling of  Opus eq. (3.10),
\begin{equation}
\hat{H}_{{\rm MB}}
\equiv\sqrt{\gamma}\sum_{i=1}^{N}
\sum_{a=x,y,z}\hat{\sigma}_{a}^\hi  \hat{B}_{a}^\hi  { ,} \label{ham4}
\end{equation}
with $\gam\ll1$ and 
where the bath operators read
\BEQ
 \hat{B}_{a}^\hi     &=&\sum_{k}\sqrt{c_k} 
 \left(  \hat{b}_{k,a}^\hi     +\hat{b}_{k,a}^{\dagger\left(  n\right)  }\right)  ,
 \label{B=s}
\EEQ
for each $i,a$ there is a large set of oscillators labeled by $k$, having a common coupling parameter $c_k$.
These bosons have the Hamiltonian
\begin{eqnarray}
\hat{H}_{{\rm B}}  &
=&\sum_{i=1}^{N}\sum_{a=x,y,z}\sum_{k}\hbar\omega_{k}\hat{b}_{k,a}
^{\dagger{}^\hi } \hat{b}_{k,a}^\hi     {,} \label{Hbaths}
\end{eqnarray}
with also the $\om_k$ identical for all $n,a$. 
The autocorrelation function of $B$ defines a bath kernel $K$ which is identical for all $i,a$, 
\BEA  
{\rm tr}_{{\rm B}}\big[  \hat{R}_{{\rm B}}(0) \hat{B}_{a}^\hi ( t)  \hat{B}_{b}^{(j)}(t^{\prime})  \big] = \delta_{i,j}\delta_{a,b}\,K(  t-t^{\prime})  {\rm  ,}
\label{Kt-s}  \qquad 
\hat{B}_{a}^\hi (t)  \equiv   e^{i\hat{H}_{{\rm B}}t}  \hat{B}_{a}^\hi      e^{-i\hat{H}_{{\rm B}}t} .
\EEA
Writing $c_k=c(\om_k)$, this leads to
\BEA  
K\left(  t\right) =  \sum
_{k}c\left(  \omega_{k}\right)  \left(  \frac{e^{i\omega_{k}t}}{e^{\beta \omega_{k}} \mmin 1}+\frac{e^{-i\omega_{k}t}}{1\mmin e^{-\beta\omega_{k}}}\right)
\equiv  \frac{1}{2\pi}\int_{-\infty}^{+\infty}{\rm d}\omega \,e^{i\omega t}\tilde K\left(  \omega\right) .
\EEA 
The kernel $\tilde K(\om)$ can be read off and expressed in the spectral density $\rho_c(\om)=\sum_k c(\om_k)\,\delta(\om-\om_k)$,
\begin{equation}
\label{Ktilde2}
\tilde{K}\left(
\omega\right)  =\frac{2\pi}{\vert\om\vert} \rho_c \left(  \left\vert \omega\right\vert \right)  
 \frac{\omega} {e^{\beta\omega}-1}{\rm  .} \
 \end{equation}
We adopt an Ohmic spectrum with Debye cutoff,
\begin{equation}
\tilde{K}\left(
\omega\right)  =\frac{e^{-\left\vert \omega \right\vert /\Gamma}}{4}
\frac{\omega }{e^{ \omega/T}-1}  ,   \label{Ktilde}
\end{equation}
where $T=1/\beta$ is the temperature of the phonon bath and $\Gamma$ the typical cutoff frequency. 
In Opus we also consider a Lorentzian (power law) cutoff, for which the statics allows  analytic results. 

With the couplings in (\ref{ham4}), (\ref{B=s}) and (\ref{Hbaths}) independent of $a$, $\hat{H}_{{\rm MB}}$ is statistically invariant under $Z_{2l+1}$. 
Combined with the invariance of  $\hat H_\rmM$ and $\hat H_\SA$, this secures an unbiased measurement.

\subsection{Evolution of the density matrix}
\label{sec2.4}

The evolution of the density matrix  of the total system is given by the Liouville--von Neumann equation. On the eigenbasis of $\hat s_z$,
its elements  $\hat R_{s\sb}$ evolve independently as  given in  eq (4.8) of Opus; it involves the apparatus spins and the bath. 
The procedure of Opus for spin $l=\half$ appears to hold for general spin-$l$ operators.

Let us  consider the time evolution of $\hat R_{s\sb}$ as given in 
eq (4.8) of Opus (we now denote $i\to s$, $j\to \sb$), where the action of the harmonic oscillator bath has been expressed in the bath kernel $K(t)$
and which involves commutators of $\hat R_{s\sb}$ with the spin operators $\hsig_a^\hi$, $a=x,y,z$;  $i=1,\cdots ,N$.

Formally, the initial state (\ref{paraM}) is a (constant)  function of the $\hsig_z^\hi$.
Also $\dot R_{s\sb}(\ti)$ is a function of them, so
 it is consistent to assume that at all $t$, $\hat R_{s\sb}$ only depends on the $\hsig_z^\hi$.
As a result,  the $a=z$ terms eq. (4.8) in Opus have vanishing commutators for any spin $l$.
Left  with the $x,y$ commutators, we define
(using the index $n$ rather than $i$ to label the  $\hsig_{x,y}$)
\BEQ
\hat\sig_\pm^\hen =\hat\sig_x^\hen  \pm i \hat\sig_y^\hen.
\EEQ
Because $\sum_{a \iss x,y} \hat\sig_a^\hen \hat O\hat\sig_a^\hen \iss \half\sum_{\alp=\pm1}
\hat\sig_\alp^\hen \hat O\hat\sig_\malp^\hen$ 
for any operator $\hat O$,   eq. (4.8) in Opus takes the form
\BEQ  \label{dRdts12} \label{dRij12}
&&  \hspace{-6mm}
\frac{{\rm d}\hat{R}_{s\sb }(t)}{{\rm d}t}=-i \hat{H}_{s}\hat{R}_{s\sb }(t)+i\hat{R}_{s\sb }(t)\hat{H}_{\sb } 
+ \, \frac{\gamma}{2} \sum_{\alp,\beta=\pm1} \sum_{n=1}^N  \int_{0}^{t} \d u \,  K(\beta u)  \hat C_{s\sb,\beta}^{(\alp,n)}(u)  ,\nn
\EEQ
where 
\BEQ \label{CspCsmgen12} 
 \hat C_{s\sb,+}^{(\alp,n)} (u) \iss \left[ e^{-iu\hat H_s} \hsig _\malp  ^\hen e^{iu\hat H_s} \hat{R}_{s\sb }(t), \, \hsig _{\alp}^\hen\right] ,\qquad   
 \hat C_{s\sb,-}^{(\alp,n)} (u) \iss \left[  \hsig _\malp  ^\hen, \, \hat{R}_{ s \sb }(t) e^{-iu\hat H_\sb} \hsig _\alp^\hen e^{iu\hat H_\sb}   \right]  ,
\EEQ
are commutators involving the Hamiltonian of M coupled to S in state $s$,  without the bath, viz.
\BEQ \label{Hshs}
\hat H_s=\hat H_\rmM +\hat H_\SA^s = N H (\hat m_{1})+N  I_s(\hat m_{1})  . \quad
\EEQ
The action of the bath is expressed in the kernel $K(\pm u)$, with the smallness of $\gam$ allowing to truncate it at its first order.
Eqs. (\ref{dRij12})  and (\ref{CspCsmgen12} ) are valid for general spin $l=\half,1,\frac{3}{2},\cdots$. 

Most importantly, the $\hat R_{s\tilde s}$ are decoupled in the separate $s,\sb$ sectors, a property for ideal measurement,
but absent in general. Examples of these non-idealities are the spin S having a nontrivial dynamics during the measurement 
and biased measurement where the Hamiltonian of the magnet and/or the bath depends on the state of S.

\subsection{Physical implementation of the model}
\label{sec2.5}

 The spin-$\half$ Curie-Weiss model for quantum measurement  \citep{allahverdyan2003curie}
was  initially conceived as a tool to understand the dominant physical aspects of idealized quantum measurements. 
It has served this task well. Let us look here at possible realizations of the model.

Curie-Weiss models are mean-field types of spin models. Their distance-independent couplings apply to a small magnetic grain.
The grain need not be very large. From studies of spin glasses and cluster glasses, it is know that ``fat spins'', clusters of 
hundreds or thousands of coherent spins,  are easily detectable \citep{mydosh1993spin}.

The Ising nature of the couplings refers to fairly anisotropic spin-spin interactions.
For spin $\half$, eq. (\ref{UN=}) expresses pair and quartet couplings between the $z$-components of the spins.
Multispin interactions are a natural result of overlap of electronic orbits; here they are approximated to do not decay with the distance
between the spins in the grain. How reasonable this approximation is, needs to be considered in each separate application.
The main feature of our modeling being a first-order phase transition in the magnet, suggests that it represents a large class of short-range  systems.
This is underlined  by the model's support of the Copenhagen postulates of collapse and Born probabilities.

For the spin-1 Curie-Weiss model, these features hold too.
But on top of this, eq. (\ref{exps2s}) produces the combination $\hat\Sigma_i\equiv \hsig_z^{(i)\,2}-{2}/{3}$, which takes the value ${1}/{3}$ for
the ``out-of-plane'' cases  $\sig_i=\pm1$, and $-2/3$ for the ``in-plane'' case $\sig_i=0$.
Separate-spin terms of the form $\sum_i D\hsig_z^{(i)\,2}$  are well known, stemming from crystal fields.
For the apparatus, the $\co_1^2$ term of eq (\ref{C2a=}) relates to the interaction $\sum_{ij} \hat\Sigma_i\hat\Sigma_j$ 
between the $\hat\Sigma_i$, so it involves the mentioned $D$-term but also the terms $\hsig_z^{(i)\,2}\hsig_z^{(j)\,2}$.
 How to implement such crystal-field-type spin-spin interactions in practice is an open question. 

Concerning numerical implementations: the \cite{{donker2018quantum}} approximation of the Curie-Weiss model can be generalized to higher spin.

\renewcommand{\thesection}{\arabic{section}}
\section{The spin $\frac{1}{2}$ case revisited}
\renewcommand{\theequation}{3.\arabic{equation}}
\setcounter{equation}{0} 
\renewcommand{\thesection}{\arabic{section}}

\label{sec3}

\subsection{Elements of the statics}
\label{sec3.1}

We set the stage by considering the spin-$\half$ situation, the original Curie-Weiss model for quantum measurement in slightly adapted 
notation\footnote{For the connection with the parameters in Opus, see ref 1.}.
The spin operators are $\hat \sig_{x,y,z}$, with $\hsig_z={\rm diag}(\half,-\half)$.
 It holds that $[\hsig_a,\hsig_b]=i\eps_{abc}\hsig_c$ and $\hsig_x^2+\hsig_y^2+\hsig_z^2=\frac{3}{4}\hat \sig _0$ with $\hat \sig_0={\rm diag}(1,1)$.

The magnet has $N$ such spins $\hsig^\hi_{x,y,z}$, $i=1,2,\cdots , N$.
They have magnetization operator
\BEQ
\hat M_1=N\hat m_1,\quad \hat m_1=\frac{1}{N}\sum_{i=1}^N \hsig_z^\hi, \qad 
\EEQ
taking eigenvalues $-\half\le m_1\le\half$.
 In the paramagnetic state, $m_1=0$.
The Hamiltonian is taken as pair and quartet interactions,
\BEQ
\label{UN=}
\hat H_\rmM =N\hat H,\qquad 
\hat H=-2J_2\hat m_1^2-4J_4\hat m_1^4 .
\EEQ
With $\hat x_\sigma=\hat N_\sigma/N$
it holds that
\BEQ
\hat m_1=\frac{1}{2}\hat x_{1/2} - \half\hat x_{-1/2 },
\quad 
\hat x_{\pm 1/2}=\frac{1}{2}\hat I \pm\hat  m_1 .
\EEQ

The spins have eigenvalues $\sig_i=\pm\half$, so that $\hat m_1$ 
has eigenvalues $m_1=\nu\sum_i\sig_1$ ranging from $-\half$ to $\half$ with steps of $2\nu$.

\subsection{The interaction Hamiltonian}
\label{sec3.2}

In order to use the magnet coupled to its bath as an apparatus for a quantum measurement,
a system--apparatus (SA) coupling is needed. According to (\ref{HIcsh}) it is chosen as a spin-spin coupling, 
\BEQ\label{HSAshalf}
\hat H_\SA=-4g\sum_{i=1}^N\hat s\hsig_z^\hi=-4gN\hat s\hat m_1 ,
\EEQ
and takes the values $H_\SA^s(m_1)=-4gsNm_1$.
The full Hamiltonian of S+A in the sector $s$ thus reads
\BEQ \label{Hs12}
\hat H_s=
 -2J_2N\hat m_1^2-4J_4N\hat m_1^4  -4gsN  \hat m_1.
\EEQ

The eigenvalues of $\hat s_z$ are $s=\pm\half$ and those of $\hsig^\hi_z$ are $\sig_i=\pm\half$, so that  $\hat m_1$
has the eigenvalues $\nu\sum_{i=1}^N\sig_i$. 
The degeneracy of a state with magnetization $m_1$ is 
\BEQ \label{Gs12}
\GN = \frac{ N!}{ N_\mhalf !\, N_\half!}
=\frac{N!}{(Nx_\mhalf)! (Nx_\half)!},
\EEQ
and entropy  $S_N=\log \GN$. At large $N$, we get the standard result for the entropy $S_N=NS$ with
\BEQ && \hspace{-10mm}
S =  \\ &&  \hspace{-10mm}
 \mmin\frac{1\mmin2m_1}{2}\log\frac{1\mmin2m_1}{2}  \mmin\frac{1\pplus2m_1}{2}\log\frac{1\pplus2m_1}{2}  . \nn
\hspace{-10mm}{} 
 \EEQ

Combining (\ref{Hs12}) and (\ref{Gs12}) the free energy in the $s$-sector reads
\BEQ \label{Fs1/2}
 F_s(m_1)=
 -2J_2Nm_1^2-4J_4Nm_1^4  -4gsNm_1-T\log \GN(m_1),
 \EEQ
 which yields for large $N$
 \BEQ
 \frac{F_s}{N}=\mmin 2J_2m_1^2 \mmin  4J_4m_1^4 \mmin 4gsm_1\mmin TS(m_1).
\EEQ

\renewcommand{\thesection}{\arabic{section}}
\subsection{Dynamics of the spin $\half$ model}
\renewcommand{\theequation}{4.\arabic{equation}}
\setcounter{equation}{0} 
\renewcommand{\thesection}{\arabic{section}}

\label{sec4}

\myskip{
\renewcommand{\thesection}{\arabic{section}}
\section{Recap: Dynamics of the spin $\half$ model}
\renewcommand{\theequation}{6.\arabic{equation}}
\setcounter{equation}{0} \setcounter{figure}{0}
\renewcommand{\thesection}{\arabic{section}}
}

\label{sec6}

At the initial time $\ti$ of the measurement, the state of the tested system S, here $\hat{\bf s}$, a spin-$\half$ operator, 
is described by its $2\times 2$ density matrix $\hat r(\ti)$ wth elements $r_{s\sb}(\ti)$ for $s,\sb=\pm\half$.
The magnet M has $N\gg1$ quantum spins-$\half$   {\boldmath{$\hat \sigma^\hi$}} ($i=1,\cdots,N$).
In each $s,\sb$ sector, S+M lie in the state $\hat R_{s\sb}(t)=\hat R_{\sb s}^\dagger(t)$, which is an operator that can be represented
by a $2^N \!\times 2^N$ matrix. At $\ti$, M is assumed to lie in the
 paramagnetic state wherein the spins are fully disordered and uncorrelated. Multiplying by the respective element of $\hat r(\ti)$ this leads to
 the elements of the initial density matrix of S+M 
 \BEQ \label{paraM}
\hat R_{s\sb}(\ti)=r_{s\sb}(\ti)\frac{\hsig_0^\heen}{2}\otimes\frac{\hsig_0^\htwee}{2}\otimes\cdots\otimes \frac{\hsig_0^\hN}{2} .
\EEQ

\subsection{Truncation for spin $\half$}
\label{sec4.1}

Dynamics of the off-diagonal elements (cat terms) was worked out in Opus. In the relevant short-time domain, the spin-spin couplings are ineffective,
so that it suffices to study independent spins coupled by the interaction Hamiltonian and to the bath. 
These elements vanish dynamically, truncating the density matrix $\hat R$ to a form diagonal on the eigenbasis of $\hat s_z$.
There is no reason to repeat that here;
for spin-1 this will be worked out in section \ref{sec5.1}.

\subsection{Registration for spin $\half$}
\label{sec4.2}
Registration of the measurement is described by the evolution of the diagonal elements of the density matrix of the full system.
 For the situation of higher spin it is instructive to reconsider and slightly reformulate the spin $\half$ situation.
 
For $\sb=s$ the Hamiltonian terms drop out of (\ref{dRij12}), hence the dynamics is a relaxation set by 
\BEQ  &&  \hspace{-8mm}
\frac{{\rm d}\hat{R}_{ss }(t)}{{\rm d}t}=
\frac{\gamma}{2} \sum_{\alp,\beta=\pm1} \sum_{n=1}^N  \int_{0}^{t} \d u \,  K(\beta u)  \hat C_{ss,\beta}^{(\alp,n)}(u) .
\label{dRii12}
\EEQ

For $l=\half$, the spin operators $\hsig_{x,y,z}$  anticommute, hence for any function $f$ of the $\hsig_z^\hi $ it holds that 
\BEQ 
\hsig_{\alp}^\hen  f(\{\hsig_z^\hi\})
= \hat f( \{(-1)^{\delta_{i,n}}   \hsig_z^\hi  \})\hsig_{\alp}^\hen  
\equiv   f^\hen ( \{ \hsig_z^\hi  \})  \hsig_{\alp}^\hen.
\EEQ 
This brings the $\hsig_\alp^\hen$ and $\hsig_\malp^\hen$ next to each other, which allows to eliminate them  using  the sum 
$\sum_{\alp=\pm1}\hsig^\hen_\alp\hsig^\hen_\malp=\hsig_0^\hen$.
With only functions of the $\hsig_z^\hi$ ($i=1,\cdots,N$) remaining,  we can go to their diagonal bases, so as to work with scalar functions of their eigenvalues $\sigi=\pm\half$, 
(see also Opus, sect. 4.4). This expresses (\ref{CspCsmgen12}) as
\BEQ \label{C+C-1/2}
&& \hspace{-9mm}
C_{s\sb,+}^\hen (u)  \equiv \sum_{\alp=\pm1}C_{s\sb,+}^{\alp\,n}(u)  
= e^{-iuH_s} e^{iuH_s^\hen } {R}^\hen_{s\sb }(t) - e^{-iuH_s^\hen} e^{iuH_s} {R}_{s\sb }(t) ,
\nn\\&& \hspace{-9mm}
C_{s\sb,-}^\hen (u) 
\equiv \sum_{\alp=\pm1}C_{s\sb,-}^{\alp\,n}(u)  
=  {R}^\hen_{s\sb }(t)e^{-iuH^\hen_\sb} e^{iuH_\sb}  - {R}_{s\sb } (t)e^{-iuH_\sb}  e^{iuH_\sb^\hen} ,  
\EEQ
where for any function $f(\{\sigi\})$, $f^\henn$ has the sign of $\sig_n$ reversed, 
\BEQ
f^\henn(\{\sig_i\})= f(\{ (-1)^{\delta_{i,n}} \sigi \}) .
\EEQ
We employed the obvious rules $(fg)^\hen=f^\hen g^\hen$ and $[f(g)]^\hen=f(g^\hen)$. The terms in (\ref {C+C-1/2}) being scalars yields
the relation $C_{s\sb,-}^\hen (u)=C_{s\sb,+}^\hen (-u)$, which allows to combine the integrals of (\ref{dRdts12}) and (\ref{CspCsmgen12}) 
as a single one from $u=-t$ to $t$.
Since $\gam\ll1$,  the typical scale of $t$, the registration time $1/\gam T$, is much larger than the bath equilibration time $1/T$.
Hence we may now take the integral over the whole real axis to arrive at the Fourier transformed kernel $\bar K(\om)$ at specific frequencies.

The next step is to reduce the $2^N\times 2^N$ matrix problem to a problem of $N$ variables, by
considering  $R_{ss}(\{\sig_i \})=R_{ss}(m_1)$ to be functions of the order parameter $m_1=\nu\sum\sigi$. 
This is formally true at $\ti$ and valid for $\dot R_{s\sb}(\ti)$, hence it remains valid in the course of time.
Denoting $P_s(m_1)$ as the probability that $R_{ss}(\{\sigi\})/r_{ss}(\ti)$ involves $m_1=\nu\sum_i\sigi$, it picks up the degeneracy number $\GN$ in (\ref{Gs12})
of realizations $\{\sigi\}$ with the same $m_1$,
\BEQ\label{Psm1}
P_s(m_1)=\GN(m_1)\frac{R_{ss}(m_1)}{r_{ss}(\ti)}.
\EEQ
To obtain the evolution of  $\dot P_s$,  we multiply (\ref{dRii12}) by $\GN(m_1)/r_{ss}(\ti)$.
At given $m_1$, one has $m_1^\hen=m_1-2\nu \sig_n$, so we can split in terms with $\sig_n=\half$ (and $-\half$), and perform the sum over $n$.
The fraction of terms that flips an up spin $\sig_n=\half$ is $x_\half(m_1)$, which multiplies $P(m_1-\nu)$;  
flipping a down spin $\sig_n=-\half$ happens with probability $x_\mhalf(m_1)$, which multiplies $P(m_1+\nu)$. 
Due to eq.  (\ref{Psm1}) these $P$'s involve the ratios
\BEQ\label{G2Gppm} &&
\frac{\GN(m_1)}{\GN(m_1-\nu)} =\frac{x_\mhalf(m_1-\nu)}{x_\half(m_1)},\quad
\quad 
\frac{\GN(m_1)}{\GN(m_1+\nu)} =\frac{x_\half(m_1+\nu)}{x_\mhalf(m_1)} ,
\EEQ
which has the effect of eliminating  the $x_{\pm\half}(m_1)$.
Introducing the operators $E_\pm$ and  $\Delta_\pm=E_\pm-1$ by 
\BEQ && 
E_\pm f(m_1)= f(m_1\pm \nu), 
\qquad \Delta_\pm f(m_1)=f(m_1\pm \nu)-f(m_1), 
\EEQ 
the evolution of $P_s$ gets condensed as
\BEQ \label{dotPs12alp} &&
\dot P_s(m_1)=\frac{\gam N}{2} 
\sum_{\alp=\pm1} \Delta_{\alp} \{ x_{\half \alp}(m_1)\tilde K[\Om_{s\,\alp}(m_1)] P_s(m_1) \}  .
\EEQ
where
\BEQ\label{OmHs12}
\hspace{-0mm}
\Om_{s\, \pm}(m_1) \iss \Delta_\mp H_s \iss H_s(m_1\mp\nu) \mmin H_s(m_1) .
\EEQ
This is now a problem for $N+1$ functions $P(m_1;t)$ subject to the normalization $\sum_{m_1}P(m_1;t)=1$.

In fig. 1, the distribution of the magnetization $m_1$ is depicted at various times. In fig. 2, this evolution is represented in a $3d$ plot.

\begin{figure}
\centerline{ \includegraphics[width=8cm]{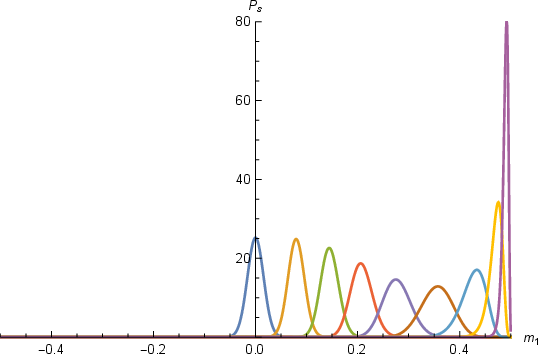}  \includegraphics[width=8cm]{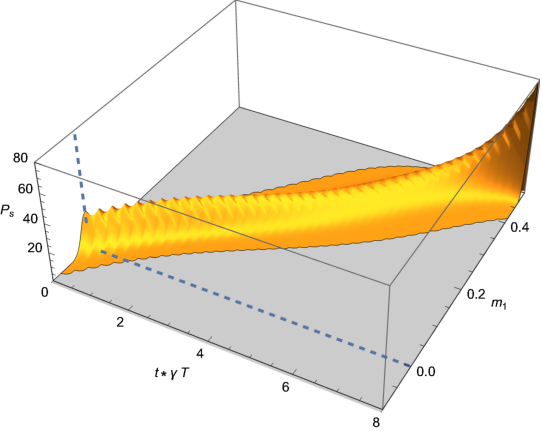}}
\caption{Left: Evolution of the magnetization distribution $P_s(m_1;t)$ for $s=+\half$ at times $0,1,\cdots,8$ in units of $1/\gam T$.
The paramagnetic state at $t=0$ is peaked around $m_1=0$; the coupling between S and A moves the peak towards $m_1=+\half$.
In doing so, it first broadens and later narrows significantly.  \label{fig1} }
\caption{Right: The case of figure \ref{fig1}   plotted in $3d$ at intervals $\Delta  t=0.2/\gam T$.} \label{fig2}
 \end{figure}

\subsection{$H$-theorem and relaxation to equilibrium}
\label{sec4.3}

The dynamical entropy of the distribution $P_s(m_1;t)=\GN(m_1)R_{ss}(\{\sigi\}  ;t )/r_{ss}(\ti)$ is defined as
\BEQ
S_s(t)& \iss & -{\rm Tr} \frac{\hat R_{ss}(t)}{r_{ss}(\ti)}\log\frac{\hat R_{ss}(t)}{r_{ss}(\ti)}  
=-\sum_{m_1} P_s(m_1;t)\log\frac{P_s(m_1; t)}{\GN(m_1)} . 
\EEQ
As in Opus, we introduce a dynamical free energy
\BEQ \label{Fdyn12} 
\Fdyns(t)= U_s(t)-TS_s(t) 
=\sum_{m_1} P_s(m_1;t)\left[H_s(m_1)+T \log \frac{P_s(m_1;t)}{\GN(m_1)}\right] , 
\EEQ
which adds the $P_s\log P_s$ term to the average of the free energy functional $F_N(m_1)=H_s(m_1)-TS_N(m_1)$.
With $\beta=1/T$, eq. (\ref{Peq3}) yields
\BEQ \label{dotFdynDeltaa}
\dot F_\dyn^s=T\sum_{m_1} \dot P_s(m_1) \log \frac{P_s(m_1)e^{\beta H_s(m_1)}}{\GN(m_1) } =
 \frac{\gam NT}{2}\sum_{\alp=\pm1}\sum_{m_1}  \Delta_\alpha [x_{\half\alp}\tilde K(\Om_{s\, \alp}) P_s ] 
 \log \frac{P_s e^{\beta H_s} }{\GN} . 
\EEQ
For general functions $f_{1,2}(m_1)$ and $\alp=\pm1$,  partial summation yields
\BEQ
\sum_{m_1} (\Delta_\alp f_1)f_2
=\sum_{m_1}f_1 (\Delta_\malp f_2) =
\hspace{-5mm} \sum_{m_1} E_\alp\left[f_1 (\Delta_\malp f_2)\right]
=-\sum_{m_1}(E_\alp f_1) (\Delta_\alp f_2). 
\EEQ
provided that the boundary terms  $f_{1,2}(m^\nu_\pm)$ at $m^\nu_\pm=\pm(1+\nu)$, vanish.
As discussed,  this holds for $P_s$, but also for the logarithm in (\ref{dotFdynDeltaa}), since 
we may insert a factor  $(1-\delta_{m_1,m^\nu}-\delta_{m_1,-m^\nu})$, which makes this explicit.
For $\alp=+1$ we now use the last expression and for $\alp=-1$ the second one, which yields,
also using (\ref{OmHs12}) and the property $\tilde K(-\om)=\tilde K(\om)e^{\beta \om}$ satisfied in eqs.  (\ref{Ktilde2}), (\ref{Ktilde}),
the result
\BEQ 
 \dotFdyns=
-\gam NT\sum_{m_1} \tilde K(\Delta_+H_s) 
 \big\{e^{\Delta_+\beta H_s} (E_+ x_\half)(E_+ P_s)
\mmin  x_{\mhalf} P_s\big\}\Delta_+ \log \frac{P_se^{\beta H_s}}{\GN}  .
\EEQ
The various $x$--factors are such that a term $\GN(m_1)x_\mhalf$ can be factored out, so as to yield
\BEQ
 \dotFdyns =  -\gam NT \sum_{m_1}\sum_{\beta=\pm1} \GN x_{\mhalf} \tilde K(\Delta_+H_s) 
  \big\{e^{\Delta_+\beta H_s}\big (E_+ \frac{P_s}{\GN}\big) \mmin 
   \frac{P_s}{\GN}\big\}\Delta_+ \log \frac{P_se^{\beta H_s}}{\GN} .
\EEQ
With $\Delta_+H_s=E_+H_s-H_s$, $\GN=\exp(S_N)$ and $\cF_s(m_1)=H_s(m_1)-TS_N(m_1)$,  this can finally be expressed as
\BEQ\label{Fdyndot}
 \dotFdyns(t)=
-\gam NT\sum_{m_1} x_{\mhalf} \tilde K(\Delta_+H_s)e^{-\beta F_s} 
\Big (\Delta_+ \frac{P_s}{e^{-\beta \cF_s}} \Big)  \Big( \Delta_+ \log \frac{P_s}{e^{-\beta \cF_s}} \Big) .
\EEQ
The last factors have the form $(x'-x)\log(x'/x)$, which is nonnegative, so that $\Fdyns$ is a 
decreasing function of time. Dynamic equilibrium occurs when these factors vanish,
which happens when the magnet has reached the thermodynamic equilibrium set by  the Gibbs state $P_s=e^{-\bet F_s}/Z_s$ and 
$\hat R_{ss}=e^{-\bet \hat H_s}/Z_s$, 
with $Z_s=\sum_{m_1}\exp(-\beta \cF_s)=\sum_{m_1}\GN(m_1)\exp(-\beta H_s)={\rm Tr}\exp(-\beta \hat H_s)$, as usual.
The dynamical free energy (\ref{Fdyn12}) indeed ends up at the thermodynamic one,
\BEQ \label{Fdyninf}
\Fdyns(\infty)=-T\log Z_s=F_s(g).
\EEQ
This constitutes an example for the apparatus going dynamically to its lowest thermodynamic state, 
the pointer state indicating the measurement outcome $s=\pm\half$.
The temporal evolution from $F_{\it dyn}(0)$ to $F_s(g)$ is depicted in fig. \ref{fig2}.

\subsection{Decoupling the apparatus }
\label{sec4.4}

Near the end of the measurement,  at a suitable time $t_\dc$, the apparatus is decoupled from the system  $g=0$;
in doing so, an energy $U_\dc=-\sum _{m_1}P_s(m_1; t_\dc)  H_\SA(m_1)$ has to be supplied to the magnet, which will then relax further
its nearby minimum of the $g=0$ situation, so as to provide a stable pointer indication with macroscopic order parameter $M_{1}=Nm_{1}$,
that can be read off.

\begin{figure}
\centerline{  \includegraphics[width=8cm]{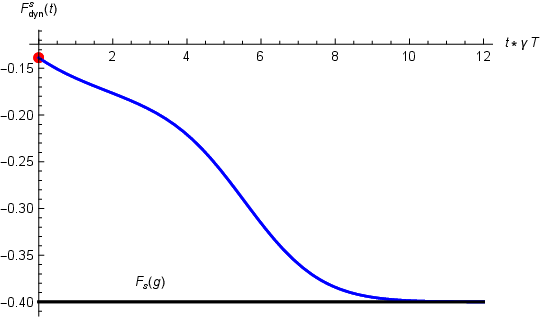}}
\caption{Evolution of the dynamical free energy 
$F_\dyn^s(t)$,  identical in both sectors $s=\pm\half$, after coupling the apparatus to a spin-$\half$ at time $t=0$.
Its approach to the Gibbs state with $F_s(g)$ (bottom line), exponential in $t$, expresses the registration of  the measurement.
} \label{fig2}
\end{figure}

\renewcommand{\thesection}{\arabic{section}}
\section{Dynamics of the spin 1 model}
\renewcommand{\theequation}{5.\arabic{equation}}
\setcounter{equation}{0} 
\renewcommand{\thesection}{\arabic{section}}
\label{sec5}
We now focus on the spin-1 case where  the tested system S is $\hat{\bf s}$, a spin-$1$ operator with $\hat s_z$ having eigenvalues $s_z=-1,0,1$.
Our magnet M has $N\gg1$ quantum spins-1  {\boldmath{$\hat \sigma^\hi$}} ($i=1,\cdots,N$).
According to eq. (\ref {mk=}) one now deals with two order parameters,
\BEQ
\hm_1=\frac{1}{N}\sum_{i=1}^N\hsig_i,\qquad 
\hm_2=\frac{1}{N}\sum_{i=1}^N\hsig_i^2. 
\EEQ 
While $\hat m_1$ is the usual magnetization in the $z$-direction, 
$\hat m_2$ is a spin-anisotropy order parameter that discriminates the sectors with eigenvalues $\sig_i=\pm1$
from the sector with eigenvalues $\sig_i=0$.

The quantity $\hat C_2$, the operator-form of eq. (\ref{C2=}), is our starting point for a permutation-invariant Hamiltonian, 
that ensures unbiased measurement.
Expanding the cosine, employing (\ref{exps2s}) for each spin $\hsig_i$, and summing over $i$, yields a polynomial in the moments $\hm_{1,2}$,
\BEQ\label{Cl=1}
\hat C_2=(1-\frac{3}{2}\hm_2)^2+\frac{3}{4}\hm_1^2 . 
\EEQ
For the Hamiltonian we take as in eq.  (\ref{UN=})
\BEQ
\hat H_N=N\hat H,\qquad \hat H=-\frac{1}{2}J_2\hat C_2-\frac{1}{4}J_4\hat C_2^2 .
\EEQ
It can be understood as to contain the single-spin term $\hm_2$, the pair couplings 
$\hm_1^2=1/N^2\sum_{ij}\hsig_i\hsig_j$ and $\hm_2^2=1/N^2\sum_{ij}\hsig_i^2\hsig_j^2$, 
the triplet couplings $\hm_1^2\hm_2$ and the quartet couplings  $\hm_1^4$, $\hm_1^2\hm_2^2$ and $\hm_2^4$.
But notice its different  conception in sec. \ref{sec2.5}.

At the initial time $\ti$ of the measurement, its state is described by its $3\times 3$ density matrix $\hat r(\ti)$ wth elements $r_{s\sb}(\ti)$ for $s,\sb=-1,0,1$.

In each $s,\sb$ sector, M lies in its state $\hat R_{s\sb}(t)=\hat R_{\sb s}^\dagger(t)$, which is an operator that can be represented
by a $3^N \!\times 3^N$ matrix. This exponential problem gets transformed in a polynomial one, a step that is exact for the considered 
mean field-type Hamiltonian.

At $\ti$, M is assumed to lie in the
 paramagnetic state wherein the spins are fully disordered and uncorrelated.
 For each spin, its state is thus $\hat\sigma_0^\hi/3$ where $\hat\sigma_0^\hi=\diag(1,1,1)$.
  Multiplying by the respective element of $\hat r(\ti)$ leads to
 the elements of the initial density matrix of S+M in the $s,\sb=0,\pm1$ sector,
 \BEQ \label{paraM}
\hat R_{s\sb}(\ti)=r_{s\sb}(\ti)\frac{\hat\sig_0^\heen}{3}\otimes\frac{\hat\sig_0^\htwee}{3} \cdots\otimes \frac{\hat\sig_0^\hN}{3} . 
\EEQ

 For general angular momentum the commutation relations $[\hat L_a,\hat L_b]=i\eps_{abc}\hat L_c$ 
 and $\hat L_x^2+\hat L_y^2+\hat L_z^2=l(\l+1)\hat I$ carry over to general spin
\BEQ\label{sigcom} \hspace{-1mm}
 [\hat\sigma_a,\hat\sigma_b] =  i\eps_{abc}\hsig_c,   \quad 
 \hsig_x^2 \pplus \hsig_y^2 \pplus \hsig_z^2  =  l(l \pplus 1)\hat \sig_0,
 \EEQ
 While we considered $l=\half$ in section \ref{sec3},  we now focus on $l=1$.

We proceed as for spin $\half$. The $a=z$ commutator in (\ref{dRdts12}) does again not contribute.
 We introduce  $\hsig_\alp=\hsig_x+i\alp\hat \sig_y$ for $\alp=\pm 1$.
From (\ref{sigcom}) it follows for general $l$ that
\BEQ \label{sigalpmalp} 
\hsig_\alp \hsig_\malp= l(l+1)\hsig_0+\alp\hsig_z-\hsig_z^2 ,\quad   \quad
(\hsig_\alp \hsig_\malp)_{\sig\sig'} =(l+1-\alp\sig)(l+\alp\sig) \delta_{\sig\sig'} . \hspace{5mm}
\EEQ
In the present case $l=1$, this has nontrivial  values
\BEQ  \label{sigalpmalpl1}
(\hsig_\alp \hsig_\malp)_{\sig\sig}= 2\delta_{\sig,\alp}+2\delta_{\sig,0} 
 ,\quad (\sig=0,\pm1), \quad
\EEQ
with (\ref{sigalpmalp}) implying that the $\sig=\malp $ term indeed drops out.
The SO(3) generators 
\BEQ\label{sigxyzl=1} && \hspace{-9mm}
\hsig_x=\frac{1}{\sqrt{2}}\begin{pmatrix} 0 & 1 & 0 \\ 1 & 0 & 1 \\ 0 & 1 & 0\end{pmatrix}   ,  \quad
\hsig_y=\frac{1}{\sqrt{2}}\begin{pmatrix} 0 & -i & 0 \\ i & 0 & -i \\ 0 & i & 0\end{pmatrix}   , \quad 
\hsig_z=\begin{pmatrix} 1 & 0 & 0 \\ 0 & 0 & 0 \\ 0 & 0 & \! \! - 1 \end{pmatrix} , 
\EEQ
allow to verify these relations. 
 Each of the   {\boldmath{$\hat \sigma^\hi$}} ($i=1,\cdots , N$) has such a presentation. 
In eq. (\ref{CspCsmgen12}) the interchange of the $\hsig_\alp^\hi$ with the $\hsig_z^\hen$ will be needed.
For $i\neq n$ they commute, while for $i=n$
\BEQ  
\hat\sigma_\alp ^{(n)}\hat\sigma_z^{(n)\,k}=(\hat\sigma_z^\hen-\alp \hsig_0^\hen)^k\hat\sigma_\alp ^{(n)}, 
\qquad \hat\sigma_z^{\hen\, k}\hat\sigma_\alp ^{(n)} =\hat\sigma_\alp ^{(n)} (\hat\sigma_z^\hen+\alp \hsig_0^\hen )^k .
\EEQ
Valid for $k=1$, induction yields this for higher $k$.
For functions of the $\{\hat\sig^\hi_z\}$, ($i=1,\cdots , N$), that can be expanded in a power series, it follows that
\BEQ && \hspace{-9mm}
\hat\sigma_\alp ^{(n)} f(\{\hat\sigma_z^\hi\}) \equiv 
f^{(n,\alp)}(\{\hat\sigma_z^\hi\} ) \hat\sigma_\alp ^{(n)} 
\qquad 
f^{(n,\alp)}(\{\hat\sigma_z^\hi\} )=
 f(\{\hat\sigma_z^\hi -\delta_{i,n}\alp\hsig_0^\hen\} ) .
\EEQ
Now the $\hsig_\pm$ can be eliminated using (\ref {sigalpmalp}), which leaves functions of only the $\hsig_z^\hi$,
with the shifts in their arguments arising as the cost for this.
As before, we can assume that $\hat R_{s\sb}(t)=R_{s\sb}(\{\hsig_z^\hi\},t)$, where  $R_{s\sb}(\{\sigi\},t)$ 
is a {\it scalar} function of the eigenvalues $\sig_i=0,\pm1$ of the $\hsig_z^\hi$. 
Valid at $\ti$, this holds for $(\d\hat R_{s\tilde s}/\d t)(\ti)$, so it remains valid in time.
Hence it is possible to go from  the matrix equations to scalar equations.
With the equality in (\ref{sigalpmalp}) applied for spin $n$, we end up with the scalar expressions
\BEQ \label{C+C-n}
&& \hspace{-10mm} 
C_{s\sb,+}^\hnpa  (u)= 
(\delta_{\sign,\malp}+\delta_{\sign,0}) e^{-iuH_s} e^{iuH_s^\hnpa} {R}_{s\sb } ^\hnpa (t)  
- (\delta_{\sign,\alp}+\delta_{\sign,0}) e^{-iuH_s^{\hnma} } e^{iuH_s} {R}_{s\sb }(t) ,
\nn\\&& \hspace{-9mm}
C_{s\sb,-}^{\hnpa } (u)= C_{s\sb,+}^\hnpa (-u) ,
\EEQ
where for any function $R_{s\sb}$ expandable in powers of the $\sig_i=0,\pm1$ ($i=1,\cdots,N$), it holds that
\BEQ &&  \hspmt
R_{s\sb}^\hnpa =R_{s\sb}(\{\sig_i\to\sig_i+\alp\delta_{i,n} \})  ,
\EEQ 
Now that all terms are scalar functions of the $\hsig_z^\hi$ it is seen that $C_{s\sb,-}^\hnpa  (u)=C_{s\sb,+}^\hnpa  (-u;H_s\to H_\sb)$.
We need no longer track the operator structure and can work with scalar functions of the  eigenvalues.

\subsection{Off-diagonal sector: truncation of Schr\"odinger cat terms}
\label{sec5.1}

In the spin $\half$ Curie Weiss model it was found that Schr\"odinger cat terms disappear by two mechanisms, 
dephasing of the magnet, possibly followed by decoherence due to the thermal bath.
Similar behavior is now investigated for spin $1$.

\subsubsection{Initial regime: Dephasing}
\label{sec5.1.1}

Truncation of the density matrix (disappearance of the cat states)
is a collective effect taking place in an early time window, 
 in which the magnet basically stays in the paramagnetic phase so that
 the mutual spin couplings $J_{2,4}$ and the coupling to the bath can be neglected. The spins of M act individually by their coupling 
 to the tested spin S, and do not get correlated yet.
In the sector where the eigenvalue of the operator $\hat s_z$ is $s$, the Hamiltonian of magnet is
\BEQ \label{HSA1} && \hspace{-9mm}
\hat H_{\SA}=\sum_n \hat H_\SA^{s\,n},\qquad 
\hat H_\SA^{s\,n} =-g\big [(1-\frac{3}{2}s^2)(\hsig_0^\hen-\frac{3}{2}\hat\sig_z^{\hen\, 2})+\frac{3}{4}s \hsig_z^\hen\big] .
\EEQ
At given $s$, this is a trace-free diagonal matrix with elements $\half g$ (twice) and $-g$,
\BEQ \label{HSAn}
&& \big(H_{\SA}^{s\,n}\big)_{\sig\tilde\sig}
=\frac{g}{2}\delta_{\sig,\tilde\sig}(1-3 \delta_{\sig,s} ) , 
\qquad  
\delta_{\sig,s}=\frac{1}{3}+(\frac{2}{3}-s^2)(1-\frac{3}{2}\sig^2)+\frac{1}{2}s \sig,  \nn
\EEQ
for $ \quad (s,\sig,\tilde\sig=0,\pm1)$. 
In this approximation, the $3^N\times 3^N$  density matrix of the magnet in each sector $s\sb$ maintains the product structure 
(\ref{paraM}) of uncorrelated spins at $t=\ti$,
\BEQ
\hat R_{s\sb}(t)=r_{s\sb}(\ti) \hat \rho_{s\sb}^\heen(t) \otimes \hat \rho_{s\sb}^{(2)}(t) \cdots  \otimes \hat \rho_{s\sb}^\heN(t), \nn\\ 
\EEQ
where, setting $\ti= 0$, for each $n$,
\BEQ\label{rhon} 
\hat \rho_{s\sb}^\hen(t)
=e^{-it\hat H_\SA^{s,n} }\frac{\hsig^\hen_0}{3}e^{i t\hat H_\SA^{\sb,n} } =\left(\hat\rho_{\sb s}^\hen(t) \right)^\dagger,\qquad  
\big(\rho_{s\sb}^\hen(t)\big)_{\sig\tilde\sig}=\frac{1}{3}\delta_{\sig,\tilde\sig} \exp\Big [\frac{3}{2}igt ( \delta_{\sig,s}-\delta_{\sig,\sb}) \Big ].
\EEQ
Diagonal elements $s=\sb$ thus essentially do not evolve in this short-time window. The off-diagonal ones imply for $s\neq\sb $
\BEQ\label{dephasing}
r_{s \sb}(t) \iss  {\rm Tr}_\rmM \hat R_{s\sb}(t) \iss r_{s\sb}(0)\left(\frac{1}{3} \pplus \frac{2}{3}\cos \frac{3}{2}gt \right)^{\hspace{-1mm} N} . 
\EEQ
For small  $t$ this decays as $r_{s\sb}(0)\exp(-t^2/\tau_{\rm dph}^2)$ with the dephasing time $\tau_{\rm dph} = 2/g\sqrt{3N}$, very short for large $N$.
The undesired recurrences at $t_n=4\pi n/3g$, where the cosine equals 1 again, can be suppressed 
by assuming that the $g\to g_n=\bar g+\delta g_n$ in (\ref {rhon}) have a small spread $\delta g_n$, see Opus, section 6.1.1.
If the thermal oscillator bath has proper parameters, it will  cause decoherence, as seen next.

\subsubsection{Second step: Decoherence}
\label{sec5.1.2}

To include the bath in (\ref{rhon}), we now make the generalized Ansatz
\BEQ 
\left(\hat \rho_{s\sb}^\hen(t)\right) _{\sig\tilde\sig} =\delta_{\sig,\tilde\sig} 
\frac{1}{3}\exp[-B_\sig(t)]
 \exp\big[ \! -i t H_\SA^{(s,n)}(\sig)+it H_\SA^{(\sb,n)}(\sig)\big] .
\EEQ
In the commutators (\ref{C+C-n}), $H_s$ now reduces to the $H_\SA^{s,n}$ of (\ref{HSAn}), and the terms are identical for all $n$.
We can neglect $B\sim\gam$ in the exponents of (\ref{dRdts12}) and find, putting $\malp\rightarrow\alp$ in the minus terms,
\BEQ
\dot B_\sig=\frac{\gam }{2} \sum_\alp \left\{
\big[K_{t>}(\Delta_\alp^\sig H_s) +K_{t<}(\Delta_\alp^\sig H_\sb) \big]  - 
\big[K_{t>}(\mmin \Delta_\alp^\sig H_s) \pplus K_{t<}( \mmin \Delta_\alp^\sig H_\sb)\big] e_{s\sb}^{\alp\sig}(t)  \right\} , \hspace{1mm} 
\EEQ
with 
\BEQ 
K_{t>}(\om) =\int_0^t\d u\,K(u)e^{-i\om u},\quad  \qquad 
K_{t<}(\om) =\int_{-t}^0\d u\,K(u)e^{-i\om u} . 
\EEQ
Here $K_{t>}(\om)=K_{t>}^\ast(\om)$ since the kernel $\tilde K(\om)$ is real valued, see the example in eq. (\ref{Ktilde}), and 
\BEQ 
\Delta_\alp^\sig H_s=H_s(\sig+\alp)-H_s(\sig) 
=\frac{3g}{2}\big [(1-\frac{3}{2}s^2)(1+2\alp\sig)-\half s\alp\big] ,
\EEQ
with a similar expression for $\Delta_\alp^\sig H_\sb$, and finally
\BEQ
e_{s\sb}^{\alp\sig}(t)= \exp[-it\,(\Delta_\alp^\sig H_s-\Delta_\alp^\sig H_\sb)] .
\EEQ
For $\sb=s$, one has $e_{s\sb}^{\alp\sig}(t)=1$. For $t\gg 1/2\pi T$, one gets, using $\tilde K(-\om)=e^{\beta\om}\tilde K(\om)$, 
\BEQ
\dot B_\sig=\frac{\gam }{2} \sum_\alp \tilde K(\Delta_\alp^\sig H_s) -\tilde K(-\Delta_\alp^\sig H_s) 
= \frac{\gam }{2} \sum_\alp (\Delta_\alp^\sig H_s)e^{-|(\Delta_\alp^\sig H_s|/\Gamma} \sim \frac{\gam}{N}, 
\EEQ
because $H_s\sim N$, $\Delta_\alp^\sig H_s\sim N^0$ and $\sum_\alp \Delta_\alp^\sig H_s\sim1/N$.
So for $s=\sb$ this confirms that hardly any dynamics takes place in this  time window;
in next subsection we show that it occurs on the longer time scale $\tau_\reg=1/\gam T$.

For off-diagonal elements $\sb\neq s$, it is seen that $e_{s\sb}^{\alp\sig}(t)$ has terms $e^{\pm3igt/2}$ and $e^{\pm3igt}$, so that
\BEQ\label{intE} 
e_{s\sb}^{\alp\sig}(t)=\sum_{j=-2,-1,1,2}c_je^{ 3ijgt /2},\qquad 
\int_0^t\d u \, e_{s\sb}^\alp(\sig;u)=\sum_{j=-2,-1,1,2}c_j \frac{e^{ 3ijgt /2}-1}{3ijg /2} .
\EEQ
The exponentials are equal to unity,  making $E_{s\sb}^\alp=1$, at the times $t_n=4\pi n/3g$, $n=1,2,\cdots$, 
encountered below eq. (\ref{dephasing}),  when appearing in the dephasing process, and thus also as times where $\dot B_\sig(t)=0$. 
To suppress recurrences like in the dephasing, we again set in each $n$-term $g\to g_n=\bar g+\delta g_n$ 
with small Gaussian distributed $\delta g_n$. For times well exceeding the coherence time $1/2\pi T$ of the bath,  
the $K_{t>}$ and $K_{t<}$ reach their finite limits, so that we have
\BEQ 
\int_0^t\d t'\, K_{t'>}(\om) E_{s\sb}^\alp(\sig;t') 
= \int_0^t\d t'\, [\,K_{t'>}(\om) -K_{\infty >}(\om) ]E_{s\sb}^\alp(\sig;t') 
+ K_{\infty>}(\om)\int_0^t\d u\,  E_{s\sb}^{\alp\sig}(u).
\EEQ
The first part is small  and the second is given in (\ref{intE}). After cancelling out its exponents by the $\delta g_n$, an imaginary part remains. 
Hence for $t\gg 1/2\pi T$, the $E_{s\sb}^{\alp\sig}$ terms can be neglected in $\Re B$. We keep
\BEQ 
\Re B_\sig(t)\approx \Re  \dot B_\sig\times t,\qquad 
\Re\dot B_\sig\approx \frac{\gam}{2} \sum_\alp  \big[\tilde K(\Delta_\alp^\sig H_s) +\tilde K(\Delta_\alp^\sig H_\sb) \big] ,
\EEQ
which is positive, so that $\vert \exp(-NB_\sig) \vert =\exp(-N\Re B_\sig)$ with $N \Re B_\sig \sim \gam N g t$ leads for large enough 
values of $N$ to a decoherence of the off-diagonal elements $r_{s\sb}(t)$  of the density matrix at the characteristic registration time
$\tau_\reg=1/\gam T$ and $\gam N g \tau_\reg\sim Ng/T$. 

Decoherence is a combined effect of the $N$ apparatus spins;  despite of it, the individual elements of $\hat R_{s\sb}$ hardly decay in this time window, 
behaving as $\exp(-\gam gt)=\exp[-(g/T)(t/\tau_\reg)]\approx 1$.

\subsection{Registration dynamics for spin 1}
\label{sec5.2}

In section 3, a difference equation was derived for the distribution of the magnetization 
of the magnet for any number $N$ of spins-$\half$.
Our aim here is to derive an analogous equation for the spin-1 case.

In the paramagnet one has the form $R_{ss}(\{\sigi\})=r_{ss}(\ti)/3^N$. Let $P_s(\vm)$ with $\vm=(m_1,m_2)$ be the probability for a state of the magnet M
characterized by the moments $m_{1,2}$. It gathers the value $R_{ss}(\vm)/r_{ss}(\ti)$ for all sequences $\{\sigi\}$ 
compatible with $m_{1,2}$, the number of which is the degeneracy factor $\GN=\exp S_N$, 
\BEQ \hspace{-3mm} 
P_s(\vm;t)=\GN(\vm)\frac{R_{ss}(\vm ;t)}{r_{ss}(\ti) }, \qquad  
\GN(\vm)=\frac{N!}{(N_{-1})!\,(N_0)!\,(N_1)!},
\quad N_\sig=x_\sig N,
\EEQ
with the $x_{\pm1}=\half(m_2\pm m_1)$ and $x_0=1-m_2$ from (\ref {xsl=1}).
The normalizations are
\BEQ
\sum_{\sig^\heen=-1}^1\cdots \sum_{\sig^\hN=-1}^1 R_{ss}(\{\sig^\hi\};t)= r_{ss}(\ti),\qquad \quad 
\sum_{m_2=0}^1 \sum_{m_1=-m_2}^{m_2}P_s(m_1,m_2;t) =1.
\EEQ

Due to the relations  (\ref{msl=1}) and  (\ref{xsl=1}) between the spin moments $m_{0,\pm1}$ and the spin fractions $x_{0,\pm1}$, the shifts in $m_{1,2}$ induce the shifts 
$N_\sig'=N_\sig+\xi_\sig$ and $x_\sig'=x_\sig+\nu\xi_\sig$,  with 
\BEQ &&   \hspace{-10mm}  
\xi_{\pm1}= \frac{1\pm\alp}{2}+\alp \sign, \qad \xi_0=-1-2 \alp\sign ,  \hspace{10mm} 
\EEQ
which are integer as they should be. The degeneracies lead  for $\sign=-\alp,0,\alp$ to the respective factors
\BEQ 
\label{Gp2Gs1}
\frac{\GN}{(\GN)'}=\frac{N_{-1}'!N_0'!N_1'!}{N_{-1}!N_0!N_1!}= \frac{x_0+\nu}{x_\malp}  \delta_{\sig_n,\malp}
+ \frac{x_\alp+\nu}{x_0} \delta_{\sign,0} + \frac{(x_{-1}
+\nu)(x_1\pplus \nu)(x_\alp \pplus 2\nu)}{x_0(x_0-\nu)(x_0-2\nu)} \delta_{\sign,\alp} ,  
\EEQ
where $N_\sig=Nx_\sig$ is used. The complicated last term is fortunately not needed, while   the denominators  of the first two will factor out.

Going to the functions $P_s$ of the moments $m_{1,2}$, we proceed as for the spin $\half$ situation. 
The $C_\pm$ terms of eq. (\ref {C+C-n}) can again be combined and performing the $u$-integrals in (\ref{dRii12})  
 leads for $t\gg 1/T$ to the kernel $\tilde K(\om)$ at  the frequencies 
\BEQ
\Om_\alp^\bet(\vm) \iss H_s(m_1 \mmin \alp\nu,m_2-\bet\nu) \mmin H_s(\vm),  \hspace{4mm}
\EEQ
for $ \alp,\bet=\pm1$.
Multiplying (\ref{dRii12}) by $G_N$ and summing over $\alp$, there results an evolution equation for the distribution $P_s$ at each discrete value of $m_{1,2}$,
\BEQ \label{Peq1} 
 && \hspace{-10mm}
\dot P_s(m_1,m_2;t)=\gam N\sum_{\alp=\pm1}   
\big\{   (x_0+\nu)   \tilde K(-\Om_{s, \malp}^{+})P_{s\,\alp}^{\,-}(m_1,m_2;t)
+ (x_\alp+\nu)\tilde K(-\Om_{s,\malp}^{\, -})P_{s\,\alp}^{+}(m_1,m_2;t) 
 \nn  \\ && \hspace{35mm}
-  \big[x_\alp \tilde K(\Om_{s\, \alp}^{+}) \pplus x_0\tilde K(\Om_{s\, \alp}^{-}) \big]P_s (m_1,m_2;t)\big\}. 
\EEQ

Let us condense notation and introduce the shift operators $E_\alp^\bet$ and $\Delta^\bet_\alp=E^\bet_\alp-1$ by their action
\BEQ 
E^\beta_\alp f(\vm)=f(m_1+\alp\nu, m_2 +\beta\nu) ,\qquad  
\Delta^\beta_\alp f(\vm)=f(m_1+\alp\nu, m_2 +\beta\nu)-f(\vm) .  \nn
\EEQ
on any $f(\met)$. They have the properties
\BEQ\label{Om2Hs} &&
E_\alp^\bet\Delta_\malp^\mbet=-\Delta_\alp^\bet,\qquad 
\Om_{s\,\alp}^{\beta}=\Delta_\malp^\mbet H_s,\qquad 
 E_\alp^\bet \Om_{s\,\alp}^\bet=-\Delta_\alp^\bet H_s=-\Om_{s,\malp}^\mbet .
\\&&
E_\alp^\bet x_\alp=x_\alp+\frac{1+\bet}{2}\nu,\quad E_\alp^\bet x_0=x_0-\bet\nu. \nn
\EEQ
Hence eq. (\ref{Peq1}) can be expressed as
\BEQ \label{Peq2} && \hspmt
\dot P_s(m_1,m_2;t)
=\gam N\sum_{\alp=\pm1}
\left( \Delta^+_\alpha [x_\alpha\tilde K(\Om_{s\,\alp}^{+}) P_s ] +\Delta^-_\alpha [x_0\tilde K(\Om_{s\,\alp}^{\,-}) P_s] \right), 
\EEQ
which has a remarkable analogy to eq. (\ref{dotPs12alp}) (and eq. (4.16) of Opus) for the spin-$\half$ 
case. By denoting $x_\alp^+=x_\alp$ and $x^-_\alp=x_0$,   this is condensed further,
\BEQ \label{Peq3}  \hspace{-2mm}
\dot P_s(m_1,m_2;t) \iss  \gam N \hspace{-2mm}
\sum_{\alp,\bet=\pm1} \hspace{-2mm}
\Delta^\beta_\alpha [x_\alp^\beta\tilde K(\Om_{s\,\alp}^{\beta}) P_s ] . \hspace{1mm}
\EEQ

\subsection{$H$-theorem and relaxation to equilibrium}
\label{sec5.3}

We now exhibit a $H$ theorem that assures relaxation of the magnet towards its Gibbs equilibrium state,
and thus a successful measurement.
The dynamical entropy of the distribution $P_s(\met ;t)=\GN(\met)R_{ss}(\{\sigi\})/r_{ss}(\ti)$ is defined as
\BEQ 
S_s(t) &=& -{\rm Tr} \frac{\hat R_{ss}(t)}{r_{ss}(\ti)}\log\frac{\hat R_{ss}(t)}{r_{ss}(\ti)} 
= -\sum_{\met }  P_s(\met ;t)\log\frac{P_s(\met ; t)}{\GN(\met)} .
\EEQ
Following Opus and eq. (\ref{Fdyn12}) above, we consider the dynamical free energy
\BEQ \label{Fdynspins} 
\Fdyns (t) = U_s(t)-TS_s(t) 
= \sum_{\met }  P_s(\met ;t)\left[H_s(\met)+T \log \frac{P_s(\met ;t)}{\GN(\met)}\right] . \nn
\EEQ
It appears to depend on $s$. The simultaneous change $s\to -s$, $m_1\to-m_1$ implies that
$F_\dyn  ^1(t)=F_{\dyn } ^{-1}(t)$ at all $t$, as happened for $s=\pm\half$ in the spin $\half$ case,
but  the $F_\dyn  ^{\pm1}(t) $ differ from $F_\dyn  ^{\, 0} (t)$, except in the thermal situations at $t=0$ and $t\to\infty$.

With $\beta=1/T$, not to be confused with the index $\beta=\pm1$, eq. (\ref{Peq3}) yields
\BEQ 
 \dotFdyns =T\sum_{\met }  \dot P_s(\met) \log \frac{P_s(\met)e^{\beta H_s(\met)}}{\GN(\met)} 
=\gam NT\sum_{\alp,\bet=\pm1}\sum_{\met }  \Delta^\beta_\alpha \big[ x_\alp^\beta\tilde K(\Om_{s\, \alp}^{\beta}) P_s \big ] 
 \log \frac{P_se^{\beta H_s}}{\GN} .
\EEQ
For general functions $f_{1,2}(\met)$ with vanishing boundary terms,  partial summation yields
\BEQ 
\sum_{\met }  (\Delta_\alpha^\beta f_1)f_2
=\sum_{\met } f_1 (\Delta_\malp^{- \beta} f_2) 
=\sum_{\met }  E_\alp^\bet\left[f_1 (\Delta_\malp^{- \beta} f_2)\right]
=-\sum_{\met } (E_\alp^\bet f_1) (\Delta_\alp^{\bet} f_2).
\EEQ
For $\alp=+1$ we use the last expression and for $\alp=-1$ the second one, with taking $\beta\to-\beta$, which yields,
also using (\ref{Om2Hs}) and the property $\tilde K(-\om)=\tilde K(\om)e^{\beta \om}$ satisfied generally in eq.  (\ref{Ktilde2}), the result
\BEQ 
 \dotFdyns =
-\gam NT\sum_{\met } \sum_{\beta=\pm1} \tilde K(\Delta_+^{\beta}H_s) 
 \big\{e^{\Delta_+^{\beta}\beta H_s} (E_+^\bet x_+^\bet)(E_+^\bet P_s)
- x_{-1}^\mbet P_s\big\}\Delta^\beta_+ \log \frac{P_se^{\beta H_s}}{\GN} .
\EEQ
The various parts are such that a term $\GN(\met)x_{-1}^\mbet$ can be factored out, to express this as
\BEQ 
 \dotFdyns \iss
\mmin \gam NT \sum_{\met } \! \sum_{\beta=\pm1} \!  \GN x_{-1}^\mbet
 \tilde K(\Delta_+^{\beta}H_s)  
 \big\{e^{\Delta_+^{\beta}\beta H_s}\big [E_+^\bet\big( \frac{P_s}{\GN}\big) \big]-
   \frac{P_s}{\GN}\big\}\Delta^\beta_+ \log \frac{P_se^{\beta H_s}}{\GN} . 
\EEQ
With $\Delta_+^\bet H_s=E_+^\bet H_s-H_s$, $\GN=\exp(S_N)$ and $\cF_s(\met)=H_s(\met)-TS_N(\met)$,  this is equal to
\BEQ\label{Fdyndot} 
 \dotFdyns (t)=
-\gam NT\sum_{\met } \sum_{\beta=\pm1} x_{-1}^\mbet \tilde K(\Delta_+^{\beta}H_s)e^{-\beta F_s}  
\times \left (\Delta_+^\bet \frac{P_s}{e^{-\beta \cF_s}}\right) \left(\Delta^\beta_+ \log \frac{P_s}{e^{-\beta \cF_s}} \right) .
\EEQ
The last factors have the form $(x'-x)\log(x'/x)$, which is nonnegative,  implying that $\Fdyns $ is a 
decreasing function of time. Dynamic equilibrium occurs when these factors vanish,
which happens when the magnet has reached thermodynamic equilibrium, that is, the Gibbs state $P_s=e^{-\bet F_s}/Z_s$ and 
$\hat R_{ss}=e^{-\bet \hat H_s}/Z_s$, 
with $Z_s=\sum_{\met } \exp(-\beta \cF_s)=\sum_{\met } \GN(\met)\exp(-\beta H_s)={\rm Tr}\exp(-\beta \hat H_s)$, as usual.
The dynamical free energy (\ref{Fdynspins}) then ends up at the thermodynamic free energy,
\BEQ \label{Fdyninf2}
\Fdyns (\infty)=-T\log Z_s,
\EEQ
which actually does not depend on $s$ due to the invariance map of the static state,
reflecting that the measurement in unbiased.
This constitutes an explicit example for the apparatus to go dynamically to its lowest thermodynamic state, 
the pointer state registering the measurement outcome.

Although the statics is identical for $s=0,\pm1$, this does not hold for the dynamics. 
While it is similar for $s=\pm1$, (to change the sign of $s=\pm1$, also change the sign of $m_1$),
this deviates from the $s=0$ dynamics. For $s=0$ all $\Om_\alp^\bet(\met)$ are finite, but for $s=\pm1$ there
are cases where $\Om_\alp^\bet(\met)$ vanishes, which leads to a slower dynamics, see fig. \ref{Fdyntspin1}.

\subsection{Numerical analysis}
\label{sec5.4}

The initial spin-1 Hamiltonian leads to a $3^N\times 3^N$ matrix problem, which is numerically hard.
For the considered mean-field type model, the formulation in terms of the order parameters $m_{1,2}$ is exact; it lowers the dimensionality considerably.
The variable  $m_2=(1/N) \sum_{i=1}^N \sigma_i^2$ can take $N+1$ values between 0 and 1.
The value of $M_2=Nm_2$ indicates that $N-M_2$ of the $\sig_i$ take the value 0, while the other $M_2$ of the $\sig_i$ are $\pm 1$.
Given this number,  $m_1=(1/N)\sum_{i=1}^N \sig_i$ can take $M_2+1$ values between $-m_2$ and $m_2$.
Accounting for conservation of total probability,  this leads to $N(N+3)/2$ dynamical variables, a polynomial problem.

(Concerning  higher spin: For spin $\frac{3}{2}$, one separates terms with $s_i=\pm\frac{3}{2}$ from those with $s_i=\pm \frac{1}{2}$;
 for spin 2 one selects terms with $s_i\iss0$, $\pm 1$ or $\pm 2$, etc.)

Eq. (\ref{Peq1}) can be solved numerically as a set of linear differential equations.
Programming it is straightforward; the vanishing of boundary terms and conservation of the total probability 
must  be verified as a check on the code.

The magnet starts in the paramagnetic initial state
\BEQ 
P_s(\met  ; 0)=\frac{1}{3^N}\GN(\met) 
\approx \frac{3^{3/2}}{2\pi N} \exp\left\{-N\left[\frac{3}{4}m_1^2 +  \left(\frac{3}{2}m_2-1\right)^2\right] \right\} . 
\EEQ
The sum of $P_s$ over $m_{1,2}$ equals unity  and, with the mesh $\Delta m_1\Delta m_2=2\nu^2$,  so does its integral.

The dynamics (\ref {Peq1}) can be solved numerically and the results are presented in upcoming figures.
We consider the parameters, with $g$ large enough,
\BEQ\label{l=1couplings} &&
N= 100
,\quad  J_2=0  ,\quad J_4=1,\quad  g=0.15,\quad
T=0.2  ,\quad \Gamma=10 .
\EEQ

\begin{figure} 
\centerline{   \includegraphics[width=8cm]{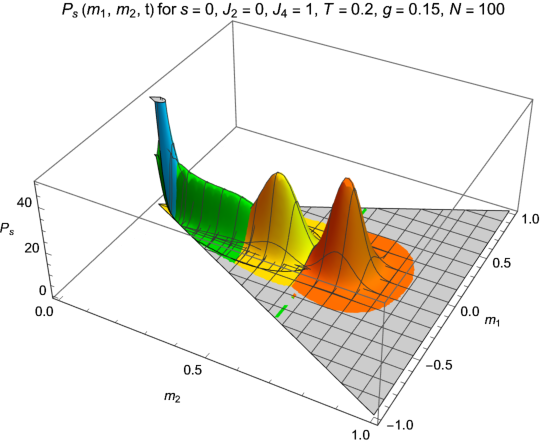}  \includegraphics[width=8cm]{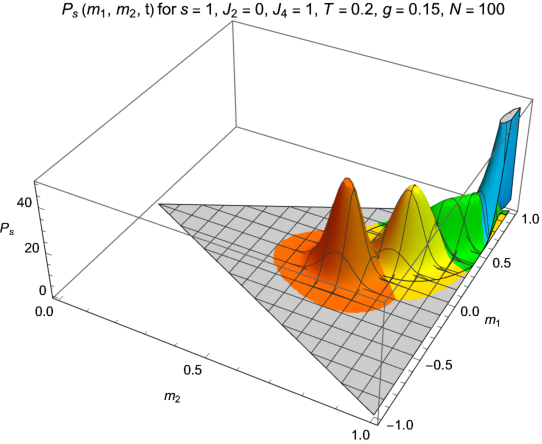} } 
\caption{Snapshots of the  distribution $P_s$
of the magnetization moments $m_{1,2}$ for registration of the spin-1 measurement. 
Left figure: $P_s\ge 10^{-3}$ data in the $s=0$ sector at times $t=(0,1,2,3)\times 2/\gam T$ from right to left.
Right figure: the $s=1$ sector at $t=(0,1,2,3)\times 5/\gam T$ from left to  right; it evolves slower.
The  parameters are listed in eq. (\ref{l=1couplings}).
} \label{figonthemove}
\end{figure}

We plot in  figs. \ref{figonthemove}a,b snapshots of $P_s/(2\nu^2)$ at four times, for $s=0$ and $s=1$.
The case $s=-1$ follows from the case $s=1$  by setting $m_1\to-m_1$.

\begin{figure}
\label{fig4}
 \centerline{  \includegraphics[width=8cm]{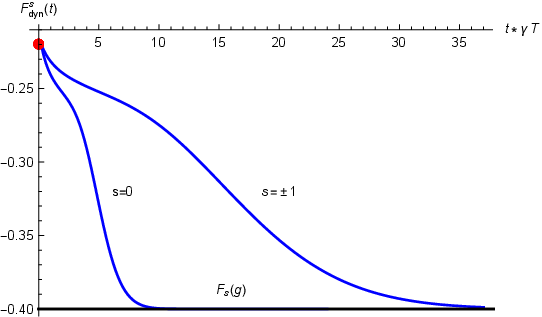} 
  \includegraphics[width=8cm]{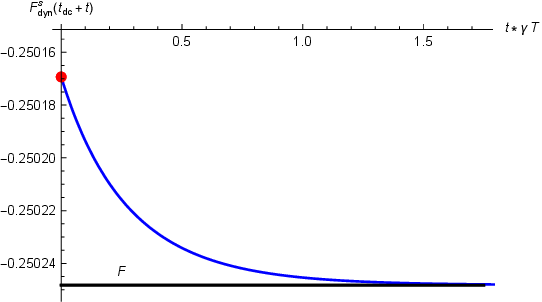} }
\caption{The spin-1 dynamical free energy $\Fdyns $ of eq.  (\ref{Fdynspins}) relaxes from its $t=0$ value
to its thermodynamic value $F_s(g)$ of (\ref{Fdyninf2}),
thereby registering the measurement. Parameters as in figs. \ref{figonthemove} a,b, and time in units of $1/\gam T$. 
The relaxation for $s=\pm1$ is slower than for $s=0$ due to the occurrence of zero frequencies.
The initial ``shoulders'' describe the initial broadenings in figs. \ref{figonthemove} a,b.
\label{Fdyntspin1} }
\caption{
After decoupling the apparatus from the system, the magnet relaxes to its nearby $g=0$ equilibrium.
If this happens  at a time $t_{\rm dc}$ where finite--$g$ equilibrium has been reached, this goes  identical in the sectors $s=0,\pm1$. 
Plotted is the dynamical free energy with parameters as figs. \ref{figonthemove} and  \ref{Fdyntspin1}, 
relaxing from its decoupled value (indicated by the dot) to its $g=0$ thermodynamic limit $F$ (lower line).
Compared to fig. \ref{Fdyntspin1}, this macroscopic energy cost is a permille effect. 
} \label{fig4}
\end{figure}

Fig. \ref{Fdyntspin1} shows the evolution of the dynamical free energy $\Fdyns(t) $.

\subsection{Decoupling of the apparatus}
\label{sec5.5}

Near the end of the measurement,  the interaction between the system and the apparatus is cut off by setting $g=0$;
in doing so at decoupling time $t_\dc$, eq. (\ref{HSA1}) expresses that an amount of energy 
\BEQ 
U_\dc=-\sum_{\met } P_s(\met;t_\dc ) H_\SA(\met) \iss\pplus gN 
\sum_{\met } P_s(\met; t_\dc)\big [(1-\frac{3}{2}s^2) (1-\frac{3}{2}m_2)+\frac{3}{4}s \, m_1\big] , 
\EEQ
has to be supplied to the magnet, leaving it with the post-decoupling free energy
\BEQ
F_\dc=\sum_\vm P_s(\vm;t_\dc)  \big(H_\rmM-T \log \GN) .
\EEQ
This post-decoupling state is not an equilibrium state;
the magnet will  now relax to the nearby minimum of the $g=0$ case.
There follows a relaxation driven by bath, with the magnet evolving under the $g=0$ Hamiltonian 
$H_\rmM (\met)$ to its Gibbs state $P_{\rm G}(\met)=\GN  \exp[-H_\rmM(\met)/T]/Z_\rmG$,  
with free energy $F_G=\sum_\vm P_G(\vm)[H_\rmM (\met)-TS_N(\met)]$.

When the decoupling time $t_\dc$ is large enough, the magnet M lies in its Gibbs state at coupling $g$,  $P_s(t_\dc) \!\sim \! \exp[-\bet H_s(\met)]$.
Due to the invariance of the $g=0$ situation, the approach to it is identical for starting in any of the sectors $s=0,\pm1$.

To compare with the dynamics that end up in one of the minima, one must restrict the Gibbs state, which has 3 degenerate minima,  to the nearby minimum.
This is achieved numerically even at moderate $N$ by discarding $\exp(- \beta H_s)$ well away from the peak of $P_s(t_\dc)$, also in $Z_\rmG$.
For $s=0$ it suffices to keep $\exp(-\beta H_s)$ for $m_2<\frac{1}{3}$;  for $s=\pm1$ by doing that for $m_1s>\frac{1}{3}$.  

The change of the state is also seen in $\langle m_2\rangle(t)=\sum_\met m_2P_s(\met;t)$. Let us consider the sector $s=0$,
where $\langle m_1\rangle=0$ at all $t$. Here the coupling $H_\SA= gN(\frac{3}{2}m_2-1)$ has the tendency to suppress $m_2$, so after decoupling  $m_2$ will relax to a larger value.
For $N\to\infty$ we get from the Gibbs states at $g$ and at $g=0$,  respectively,
\BEQ  \hspace{-3mm}
\langle m_2(0)\rangle \iss 3.63 \, 10^{-4}  , \qad 
\langle m_2(\infty)\rangle \iss 11.5.\, 10^{-4}  . \qad 
\EEQ
The full time behavior for $N=100$ and couplings as in eq. (\ref{l=1couplings}), is presented in fig. \ref{fig4}, with the finite-$N$ values increasing from
$\langle m_2(0)\rangle=9.975\, 10^{-4}$ to $\langle m_2(\infty)\rangle=12.69\,10^{-4}$.

The relaxation in the sectors $s=\pm1$  follow immediately from this. The map (\ref{m12p})  yields
\BEQ \label{m12p} 
\langle m_{1}\rangle_{s=\pm1}=\pm\big(1-\frac{3}{2}\langle m_2\rangle_{s=0}\big),   \qquad 
\langle m_2\rangle_{s=\pm1} =1-\half\langle m_2\rangle_{s=0}.
\EEQ


\subsection{Energy cost of quantum measurement}
\label{sec5.6}

The Copenhagen postulates obscure one of the facts of life in a laboratory: a firm cost for the energy needed to keep the setup running.
In this work, we consider two intrinsic costs. In previous subsection, we established the cost for decoupling the apparatus from the system.
Here we consider resetting the magnet  for another run.
It has to be set  from its stable state 
back to its metastable state.
Being related to the magnet, both costs are macroscopic.

Our initial state, the paramagnet (pm), has zero magnetic energy and maximal entropy
\BEQ
F_\pama=-NT\log 3,
\EEQ
The energy needed to reset the Gibbs state of the magnet to the paramagnetic one is 
\BEQ
U_{\rm reset}=F_\pama-F_G 
=-\sum_\vm P_G(\vm)[H_\rmM-T\log (\GN/ 3^N)].
\EEQ
It is evidently macroscopic.
The condition that $U_{\rm reset}$ is positive, was identified in Opus and in Models
as the condition that the initial paramagnetic state is metastable but not stable.

\section{Summary}
\label{sec6}
\label{summ}

This paper deals with the dynamics of an ideal quantum measurement of the $z$-component of a spin 1. 
The statics for this task has been worked out recently in our 
 ``Models'' paper \citep{nieuwenhuizen2022models}; it generalizes to any spin $l>\half$, 
the Curie-Weiss model to measure a spin $\half$; the latter was considered in great detail in ``Opus'' \citep{allahverdyan2013understanding}. 
Here we first reformulate the dynamics of the known case for spin $\half$ and work out some further properties.
The resulting formalism  is suitable as basis for models to measure any higher spin.

The dynamics of measurement in the spin-1 case is analyzed in detail. 
Off-diagonal elements of the density matrix (`cat states'') are shown to decay very fast (``truncation of the density matrix'') due to dephasing, possibly followed by decoherence.

The evolution of the diagonal elements of the density matrix  is expressed in coupled first order differential equations for the distribution 
of two magnetization-type order parameters, $m_{1,2}$. The approach to a Gibbs equilibrium is certified by demonstrating a $H$-theorem.
The resulting scheme is numerically a polynomial problem. For the present power of laptops,  these are easily solved for an apparatus consisting of a few hundreds of spins.
The evolution of the probability density is evaluated and the $H$-theorem is verified. 
The macroscopic energy costs for decoupling the apparatus from the spin, and for resetting the apparatus from its stable state
to its metastable state for use in next run of the measurement, are qauntified.

For general spin $l$ this method simplifies the numerically hard problem of dimension $(2l+1)^{2N}-1$ by a polynomial problem 
of order $N^{2l}$ for its $2l$ order parameters. For more complicated models of the apparatus, it will likewise pay off to focus 
on the order parameter of the dynamical phase transition of the pointer that achieves  the registration of the measurement. 
 The fact that phase transition in the magnet is of first order underlines that our mean field-type models, although of mathematical convenience,
are not essential for the fundamental description of quantum measurements.


 \renewcommand{\thesection}{\arabic{section}}
 \section{References} 
 \label{refs}

\end{document}